\documentclass[aps,prd,amsmath,amssymb,twocolumn,superscriptaddress,preprintnumbers,nofootinbib]{revtex4-1}
\pdfoutput=1
\usepackage{graphicx}
\usepackage{amsthm}
\usepackage{amsmath}
\usepackage{graphicx,subfigure}% Include figure files
\usepackage{dcolumn}% Align table columns on decimal point
\usepackage{bm}% bold math
\usepackage{slashed}
\usepackage{calligra}
\DeclareMathAlphabet{\mathcalligra}{T1}{calligra}{m}{n}
\DeclareFontShape{T1}{calligra}{m}{n}{<->s*[2.5]callig15}{}

\newcommand{\be}{\begin{eqnarray}}
\newcommand{\ee}{\end{eqnarray}}
\newcommand{\bea}{\begin{eqnarray}}
\newcommand{\eea}{\end{eqnarray}}

\newcommand{\keV}{{~\rm keV}}

\newcommand{\GeV}{{~\rm GeV}}

\newcommand{\mX}{m_{_\chi} }
\newcommand{\mXp}{m_{_{\chi^*}} }
\newcommand{\exE}{\delta}
\newcommand{\thetaX}{\theta_{_\chi}}

\newcommand{\mN}{m_{_N}}

\newcommand{\mNX}{M_{{\rm N}\chi}}

\newcommand{\muX}{\mu_{_\chi}}
\newcommand{\muN}{\mu_{_N}}

\newcommand{\rhoX}{\rho_{\scriptstyle \chi} }
\newcommand{\ER}{E_{_R}}
\newcommand{\EX}{E_{_\chi}}

\newcommand{\sDD}{\sigma_{_{\rm  DD}}}
\newcommand{\sDZ}{\sigma_{_{\rm  DZ}}}

\newcommand{\vbar}{\bar{v}} % the most probable speed
\newcommand{\vesc}{v_{\rm e}} % the escape velocity
\newcommand{\vobs}{v_{\rm o}} % the minimum velocity
\newcommand{\vmin}{v_{\rm m}} % the minimum velocity
\newcommand{\vmax}{v_{\rm max}} % the minimum velocity
\newcommand{\targetDensity}{n_{_{\rm Pb}}}

\newcommand{\rDet}{\rho}

\begin{document}

\title{Dark matter detection in two easy steps}
\author{Maxim Pospelov}
\email{mpospelov@perimeterinstitute.ca}
\affiliation{Department of Physics \& Astronomy, University of Victoria, Victoria, BC, V8P 5C2, Canada}
\affiliation{Perimeter Institute for Theoretical Physics 31 Caroline St. N, Waterloo, Ontario, Canada N2L 2Y5.}
\author{Neal Weiner}
\email{weiner@physics.nyu.edu}
\affiliation{Center for Cosmology and Particle Physics, Department of Physics, New York University, New York, NY 10003}
\author{Itay Yavin}
\email{iyavin@perimeterinstitute.ca}
\affiliation{Department of Physics \& Astronomy, McMaster University 1280 Main St. W. Hamilton, Ontario, Canada, L8S 4L8.}
\affiliation{Perimeter Institute for Theoretical Physics 31 Caroline St. N, Waterloo, Ontario, Canada N2L 2Y5.}

%\date{\today}

\begin{abstract}

Multi-component dark matter particles may have a more intricate direct detection signal than 
simple elastic scattering on nuclei. In a broad class of well-motivated models the inelastic excitation of dark matter particles 
is followed by de-excitation via $\gamma$-decay. In experiments with fine energy resolution, such as many $0\nu 2\beta$ decay experiments, this motivates a highly model-independent search for the \textsl{sidereal daily modulation} of an unexpected $\gamma$ line. Such a signal arises from two-step WIMP interaction: the WIMP is first excited in the lead shielding and subsequently decays back to the ground state via the emission of a 
monochromatic $\gamma$ within the detector volume. We explore this idea in detail by considering the model of magnetic inelastic WIMPs, and take a sequence of CUORE-type detectors as an example. We find that under reasonable assumptions about  detector performance it is possible to efficiently explore mass splittings of up to few hundreds of keV for a WIMP of weak-scale  mass and transitional magnetic moments. The modulation can be cheaply and easily enhanced by the presence of additional asymmetric lead shielding. We devise a toy simulation to show that a specially designed asymmetric shielding may result in up to 30\% diurnal modulations of the two-step WIMP signal, leading to additional strong gains in sensitivity. 

\end{abstract}

\maketitle

%%%%%%%%%%%%%%%%
% Introduction
%%%%%%%%%%%%%%%%
\section{Introduction}
\label{sec:intro}

Searches for Dark Matter (DM) in underground laboratories over the past two decades have concentrated, for good reasons, on the possibility of direct detection
of weakly interacting massive particles (WIMPs). Such experiments look for the energy associated with the nuclear recoil, $\ER$, in collisions of dark matter with the nuclei of the target material. The experimental observable is then the rate of events as a function of the nuclear recoil energy, $d R/d\ER$. This quantity is connected to the microscopic cross-section describing the collision of WIMPs with the nucleus, naturally classified to be either of spin-independent or spin-dependent elastic scattering type~(see e.g.~\cite{Lewin:1995rx} for a review).

Besides a rapid expansion of the direct detection program, the recent years have been marked by growing understanding that the DM sector may not 
necessarily be the most minimal - indeed, it may contain multiple, sometimes degenerate, WIMP states, leading to a different morphology of the direct detection signal. 
For example, as was emphasized in the original work on inelastic DM (iDM) of Tucker-Smith and Weiner~\cite{TuckerSmith:2001hy}, inelastic transitions
to excited WIMP states drastically modify the characteristics of WIMP-nucleus collisions. In particular, as the inelastic threshold increases with respect to the kinetic energy of WIMPs in the galactic halo, the scattering rate rapidly diminishes. Experiments with different targets may then be sensitive to vastly different cross-sections.

Moreover, the structure of the multi-component WIMP sector can be such that direct detection of dark matter is more advantageous with the 
use of underground detectors designed for other purposes. For example, the existence of the charged excitation of 
WIMPs \cite{Pospelov:2008qx} can be very efficiently searched in the nucleus-WIMP recombination channel 
with the  $0\nu 2\beta$ decay experiments, that  as a rule are more sensitive to this type of models than the purposely built direct DM detection 
experiments \cite{An:2012bs}. In this work, we point out that  $0\nu 2\beta$ decay experiments, at least for some models of DM, can be a "tool of choice" for detecting the 
neutral excitations of WIMPs. Such experiments can therefore serve to complement traditional direct-detection efforts and help ensure that no stone is left unturned in the search for DM. 

\begin{figure}
\begin{center}
\includegraphics[width=0.48 \textwidth]{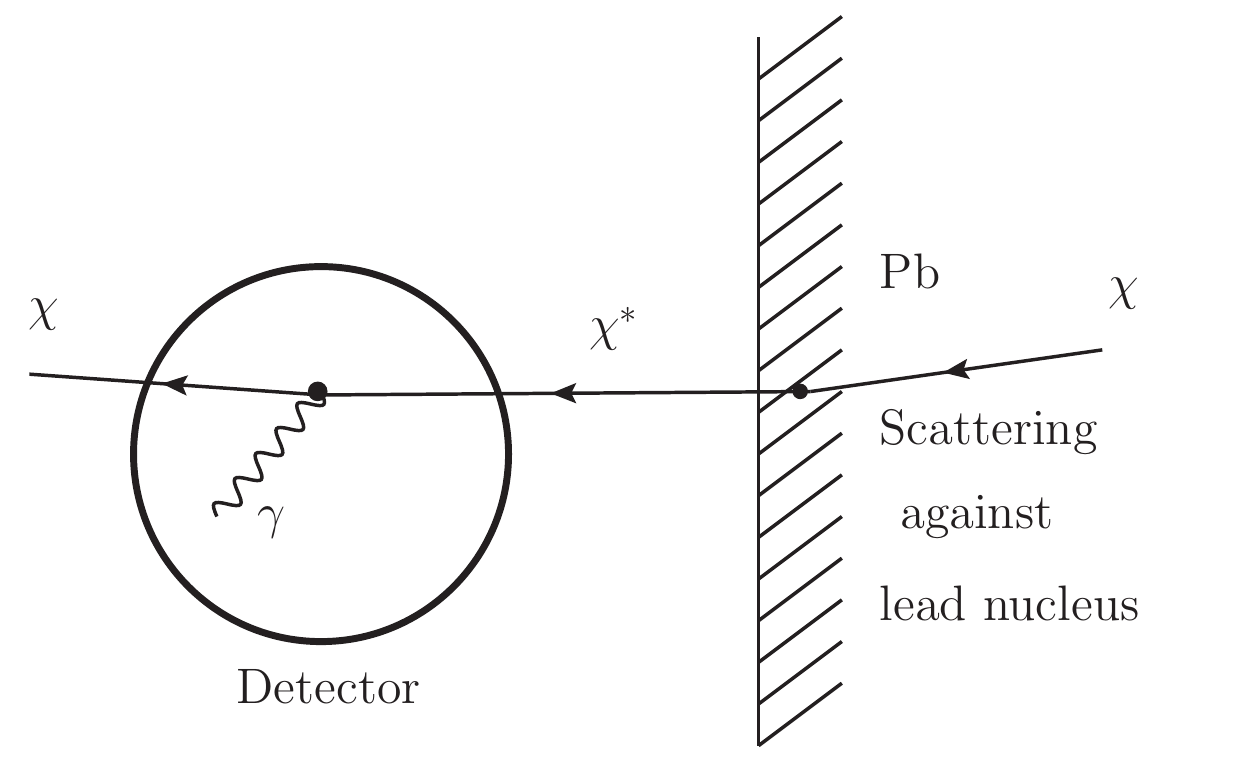}
\end{center}
\caption{A schematic of the search proposal. An incoming WIMP $\chi$ scatters in the lead shield (Pb) into the excited state $\chi^*$. The excited state then travels some distance before decaying spontaneously inside the detector into a the ground state $\chi$ and a photon. We emphasize that the decay occurs spontaneously and is independent of the detector material. }
\label{fig:setup}
\end{figure}

The idea of an inelastic transition being a dominant scattering channel is well-motivated both from known physical systems such as atoms and nuclei\footnote{Indeed, several recent proposals used the analogy with atoms~\cite{Kaplan:2009de,Kaplan:2011yj} and nuclei~\cite{Alves:2009nf} to build explicit models of DM. } as well as from the underlying theoretical structure. Furthermore, a general argument can be 
made~\cite{TuckerSmith:2001hy}, showing that pseudo-Dirac WIMP fermions allow for more freedom in the 
choice of the coupling to the SM compared to the individual Majorana components. Specifically, vector-current interactions among a multiplet of Majorana fermions are necessarily off-diagonal~\cite{ArkaniHamed:2008qn}. As in the case of atoms and nuclei, the inelastic transition can be mediated by a single-photon exchange, a possibility that was recently explored in the magnetic iDM proposal (MiDM)~\cite{Chang:2010en}. Aside from its possible connection to recent astrophysical anomalies~\cite{Weiner:2012cb,Cline:2012bz,Weiner:2012gm}, the MiDM model forms a part of a more general line of inquiry into the effects of the electromagnetic form-factors of DM in direct and indirect searches~\cite{Pospelov:2000bq,Sigurdson:2004zp}. The experimentally interesting parts of parameter space to explore are DM masses on the order of the electroweak scale,~$\mX\sim 10-10^3~{\rm GeV/c^2}$ or above, and an inelasticity of order the kinetic energy of the reduced DM-nucleus system, $\sim\mathcal{O}(100~\keV)$. A much larger inelastic threshold would make such collisions kinematically inaccessible in an on-shell process. 

Aside from modifying the kinematics of collisions, the inelasticity implies the existence of an excited state of DM that leaves the collision site. While the probability of it itself interacting again in the detector is entirely negligible, it may decay back to the ground state through the emission of an x-ray photon which can be easily detected. Importantly, since DM is much heavier than the transition energy, the line will be essentially monochromatic. Therefore, it is possible to search for such a monochromatic line independently of the original interaction involving the nuclear recoil as was first suggested in ref.~\cite{Feldstein:2010su} and later in~\cite{Chang:2010en}. In fact, the lead shield around these detectors forms an ideal target for such collisions as we propose in this paper. As shown in Fig.~\ref{fig:setup}, an incident WIMP $\chi$ can scatter in the shield and transition into an excited state, $\chi^*$, which spontaneously decays at some later time inside the detector. While the nuclear recoil energy lost by the WIMP in the lead shield is of course unmonitored, the energy emitted by the de-excitation can be observed directly by the detector.

Such a search can be easily undertaken without any specific description of the underlying model or physics that gives rise to it. Moreover, because the signal arises from the decay of particles after scattering, it will naturally modulate with the period of a sidereal day, allowing a definitive separation from background processes. Even without the following discussion of this paper, searches for the diurnal modulation of unidentified lines is a natural - if speculative - search for new physics within these experiments.

It is the purpose of this paper to spell out the details of this search proposal and provide the theoretical framework within which one can interpret the experimental results. It is a novel way to look for dark matter, but which, importantly, can simply utilize the existing technology and well-developed experimental apparatus. In the next section we survey some WIMP models that possess $\chi-\chi^*$ transitions. In section~\ref{sec:scattering_rate}, we
 calculate the scattering rates due to the inelastic up-scattering in the lead shield, and in section~\ref{sec:decay_rate} we work out the expected rate of $\gamma$ de-excitation in the detector. In section~\ref{sec:monte_carlo} we present the results of a simulation of such two-step detection process, and deduce the region of parameter space of MiDM that can be explored with CUORE-type detectors. The simulation also allows us to address the question of daily modulations of the signal and propose a simple non-invasive modifications to the $0\nu 2\beta$ experiments that would enhance such modulations. Section~\ref{sec:conclusions} contains our conclusions.

\section{Models}
\label{sec:models}

\subsection{Heuristics}
Before discussing a specific model, there are some heuristic features of any model which can be probed with this method that can be easily understood. The kinetic energy available for the collision in the center of mass frame is $\tfrac{1}{2} \mNX v^2$, where $\mNX$ is the reduced mass of the WIMP-nucleus system and $v$ is the incoming WIMP velocity. For very heavy WIMPs this is approximately $\tfrac{1}{2}\mN v^2$ and for very light ones it is $\tfrac{1}{2}\mX v^2$. If we take the escape velocity in the solar neighborhood to be $\vesc = 544$ km/s and the Sun's rotation speed with respect to the halo to be $220$ km/s then the maximum incoming speed is $\approx 2.5 \times 10^{-3}$ in units of the speed of light. In the case of a very heavy WIMP scattering against a lead target the maximum inelastic threshold accessible is therefore $\lesssim 650$ keV. In the case of a very light WIMP the inelasticity cannot exceed $\approx 3\keV \times (\mX/\GeV)$. Thus, only excited states no heavier than a few hundred keV of the WIMP itself are kinematically accessible in the laboratory. 

Given this range of accessible excitation energies, the decay back to the ground state must be through the emission of a single photon, a graviton, or two neutrinos\footnote{This assumes no new light particles associated with dark matter, an exciting possibility all by itself, but one which we eschew in this paper. It also assumes that the excited state actually can decay, see refs.~\cite{Finkbeiner:2009mi,Batell:2009vb} for the phenomenology of metastable states of dark matter.}. A decay into a graviton or neutrinos would be unobservable in the laboratory. The photon is therefore the only portal through which such de-excitations can be observed. Moreover, the mass of the WIMP cannot be much lighter than a GeV or else the de-excitation energy falls below a keV and is much too soft to be observed. Therefore, since the WIMP is always much heavier than the de-excitation energy and moves with $v/c \sim 10^{-3}$, the resulting photon is essentially monochromatic - a feature which greatly aids its detection. 

An exception to the above reasoning is found when the WIMP is partly made of a long-lived metastable state as discussed below. In that case the de-excitation energy may be somewhat larger than the excitation energy, possibly in the few MeV range. Moreover, decays into electron-positron pairs may be energetically possible. We emphasize that these possibilities can in fact be searched for with the same $0\nu 2\beta$ decay experiments generally envisioned in this paper. In fact, de-excitation energies in the MeV range enjoy a much reduced background and even decays into electron-positron pairs can be searched for by utilizing existing techniques to reconstruct background radiative peaks. 

\subsection{Magnetic Inelastic Dark Matter}

The considerations above provide a good guide for the range of WIMP mass and excitation energy that can be explored with the present proposal. However, they are insufficient to determine the precise rates and lifetimes expected. For concreteness, in this paper we will concentrate on the MiDM scenario where the WIMP scatters against matter through a magnetic dipole transition to an excited state. We comment on a more general model in the next subsection and discuss the associated parameter space of interest. The interaction Lagrangian in the case of MiDM is
\be
\label{eqn:MiDMInteraction}
\mathcal{L}_{\rm dipole}= \left(\frac{\muX}{2}\right)\bar \chi^* \sigma_{\mu\nu} F^{\mu\nu} \chi + \mathrm{h.c.},
\ee
where throughout we use natural units with $\alpha = e^2/4\pi$, and here $\muX$ is the magnetic dipole strength, $F^{\mu\nu}$ is the electromagnetic field-strength tensor, and $\sigma_{\mu\nu} = i[\gamma_\mu,\gamma_\nu]/2$ is the commutator of two Dirac matrices. Since WIMPs are heavy particles, the natural measure of their magnetic moment is the nuclear magneton, that we define as $\mu_N=\sqrt{4\pi\alpha}/(2m_p)$. 
The WIMP $\chi$ and the excited state $\chi^*$ form a pseudo-Dirac pair with a small mass splitting $\mXp - \mX = \exE$. The interaction rate in lead is determined by the differential cross-section~\cite{Chang:2010en},
\be
\label{eqn:QEDformula}
\frac{d \sigma}{d \ER} = \frac{d\sDD}{d \ER}+\frac{d \sDZ}{d \ER},
\ee
which is the sum of two contributions: the scattering due to dipole-dipole interaction between the nucleus and the WIMP,
\begin{eqnarray}
\nonumber
\frac{d \sDD}{d \ER}&=& \frac{16 \pi \alpha^2  \mN}{v^2} \left(\frac{\mu_{\rm Nucleus}}{e}\right)^2 \left(\frac{\mu_\chi}{e}\right)^2 \\[-.2cm]
\\[-.2cm]
\nonumber
 &\times& \left(\frac{S_\chi+1}{3 S_\chi}\right) \left(\frac{S_N+1}{3 S_N }\right)F_D^2[\ER];
\end{eqnarray}
and the scattering of the WIMP's dipole against the nucleus' charge, 
\begin{eqnarray}
\label{eqn:sigmaDZ}
\nonumber
\frac{d \sDZ}{d \ER}&=& \frac{4\pi Z^2 \alpha^2}{\ER}\left(\frac{\mu_\chi}{e}\right)^2 \bigg[1-\frac{E_R}{v^2}\left(\frac{1}{2m_N}+\frac{1}{m_\chi}\right)\\[-.15cm]
\\[-.15cm]
\nonumber
&-&\frac{\delta}{v^2}\left(\frac{1}{M_{\scriptstyle{N\chi}}}+\frac{\delta}{2m_N E_R}\right)\bigg]
\left(\frac{S_\chi+1}{3S_\chi}\right) F^2[\ER].
\end{eqnarray}
Here, $\ER$ is the recoil energy of the nucleus in the lab frame, $v$ is the velocity of the incoming WIMP in that same frame, $Z$ is the atomic number of the nucleus in question, $\mN$ and $\mu_{\rm Nucleus}$ are its mass and magnetic dipole moment,  respectively, $\mNX$ is the reduced mass of the WIMP-nucleus system, and $S_\chi=1/2$ and $S_{\rm N}$ are the spins of the WIMP and nucleus, respectively. Finally, $F^2[\ER]$ is the usual 
%magnetic   - changed by MP
electric 
form-factor and $F_D^2[\ER]$ is the nuclear magnetic dipole form-factor.

The experimental observable relevant to the standard direct detection experiments, the rate of nuclear recoil events, is then obtained by averaging of the halo velocity distribution,
\be
\label{eqn:dRdER_formula}
\frac{dR}{d\ER} &=& {\rm N_{T}} \frac{\rhoX}{\mX}\left \langle v \frac{d\sigma}{d\ER}\right \rangle \\\nonumber&=& {\rm N_{T}} \frac{\rhoX}{\mX} \int d^3 v f( \vec{v}+\vec{\vobs}) v \frac{d\sigma}{d\ER}.
\ee
Here ${\rm N_{T}}$ is the number of target nuclei per material mass and $\rhoX \approx 0.3 \GeV/{\rm cm}^3$ is the mass density of DM near the Earth, $\vec{v}$ is the velocity of the WIMP in the lab frame, $\vec{\vobs}$ is the velocity of the local observer with respect to the halo, and $f(v)$ is the WIMP velocity distribution in the halo. The differential rate, Eq.~(\ref{eqn:dRdER_formula}), can be used to determine the observability of this scenario in the various direct-detection experiments. 

The original work of ref.~\cite{Chang:2010en} on the MiDM scenario was motivated by the discrepancy between the detection claims of the DAMA collaboration~\cite{Bernabei:2010mq} and the strong exclusions coming from other searches such as CDMS~\cite{Akerib:2005kh,Ahmed:2009zw,Kopp:2009qt} and XENON~\cite{Angle:2007uj,Collaboration:2009xb}. It was found that this discrepancy was reconciled in the MiDM scenario resulting in a much larger rate in NaI crystals due to the large magnetic moment of iodine as compared with the other experiments. A WIMP in the mass range of $\mX \sim 70 - 300$ GeV with an inelasticity in the range of $\exE \sim 100 - 150$ keV and a magnetic moment of $\muX \sim 1 - 3\times 10^{-3}\mu_{\rm N}$ could explain the DAMA observation while staying consistent with the null results of other experiments.  Since then, the new XENON100 results~\cite{Aprile:2011hi, Aprile:2012nq}, have placed this scenario under considerable tension since about 20-30 events are expected in the energy region 30-50 keV and only a handful were observed. Unfortunately, it is difficult to ascertain how severe is the tension for various reasons: there are large uncertainties on the magnetic dipole form-factor that determines the rate in DAMA; the WIMP velocity distribution in the halo is not empirically known; the expected energy range of the events is very different than the usual energy range expected in the case of elastic scattering where the experiment is optimized; maybe most importantly, the rapid decay of the excited state in the detector into a monochromatic x-ray (see Eq.~(\ref{eqn:lifetime} below) will often result in either a double-scatter event or a higher light yield, both of which would be vetoed in a search for single-scatter nucleus-WIMP interactions. Moreover, going beyond the DAMA results, it is interesting to explore the broader parameter space associated with the MiDM scenario. 

In contrast, the experiments we envision in the current proposal can decisively explore the parameter space relevant for the DAMA results and beyond as well as search for a more general model as the one we discuss in the following subsection. 

\subsection{Dark Forces and Magnetic Transitions}

The possibility that dark matter might be a part of a larger dark sector with new forces has been widely discussed in the literature in the recent years~\cite{Finkbeiner:2007kk, Pospelov:2007mp, ArkaniHamed:2008qn, Pospelov:2008jd}. In this case, the WIMP is charged under some new vector boson(s) which is mixed with the Standard Model through its kinetic term,
\be
\label{eqn:darkLag}
\nonumber
\mathcal{L}_{\rm dark~force} &=& i\bar{\chi}\bar{\sigma}^\mu\partial_\mu \chi -\mX \bar{\chi}\chi + i\bar{\chi^*}\bar{\sigma}^\mu\partial_\mu \chi^* + \mXp \bar{\chi}^*\chi^* \\ \nonumber
&~&\\ \nonumber
&+&g_{_V}\bar{\chi}^*\bar{\sigma}_\mu V^\mu \chi+ g_{_V}\bar{\chi} \bar{\sigma}_\mu V^\mu \chi^* \\ \nonumber
&~&\\ \nonumber
&-&\tfrac{1}{4} V_{\mu\nu}V^{\mu\nu} +   \tfrac{1}{2}m_{_V}^2 V_\mu V^\mu \\
&+& \frac{\epsilon}{2\cos\theta_{_W}}V_{\mu\nu} B^{\mu\nu}
\ee
where $V_\mu$ is a new massive vector-boson, $V_{\mu\nu} = \partial_\mu V_\nu - \partial_\nu V_\mu$ is its field strength, $m_{_V}$ is its mass, and $g_{_V}$ is the gauge coupling. Here, $B_{\mu\nu}$ is the field-strength of the Standard Model hypercharge gauge field and $\theta_{_W}$ is the weak mixing angle. As is well-known~\cite{Holdom:1985ag}, when the kinetic term of the vector-boson is diagonalized the mixing with hypercharge leads to Standard Model matter being charged under the new force. Thus, it leads to an interaction between the WIMP $\chi$ and normal mater that can be searched for, among other things, in direct-detection experiments (see refs.~\cite{Pospelov:2008zw, Hook:2010tw} for a general discussion of the phenomenology). The differential cross-section in this case is given by,
\be
\frac{d\sigma}{d\ER} = \frac{16\pi Z^2 \alpha \alpha_{_V} \epsilon^2 \mN }{2 m_{_V}^2 v^2}
\ee

This model can naturally be extended to non-abelian groups and more complicated dark sectors with multiplet components~\cite{ArkaniHamed:2008qn,Baumgart:2009tn,Cheung:2009qd,Chen:2009ab,Cline:2010kv}.

As was pointed out in ref.~\cite{Finkbeiner:2009mi,Batell:2009vb}, when the mass difference $\delta = \mXp - \mX$ is below the electron-positron pair creation threshold $< 2m_e$, the excited state $\chi^*$ in such theories is generally very long-lived, easily stable on cosmological time scales. The presence of such a long-lived state is both boon and bane for phenomenology and it is beyond the scope of the present work to explore its consequences. However, we note that the magnetic transition discussed in the previous subsection, Eq.~(\ref{eqn:MiDMInteraction}), leads to a rapid decay of the excited state. We can thus consider the general model, 
\be
\mathcal{L}_{\rm dark} = \mathcal{L}_{\rm dark~force} + \mathcal{L}_{\rm dipole},
\ee
containing interactions with normal matter through both a magnetic dipole moment as well as new dark forces. In such hybrid models the scattering rate in lead may be dominated by the dark force exchange while the lifetime of the excited state may be dominated by the magnetic dipole transition. 

\begin{figure}[t]
\begin{center}
\includegraphics[width=0.45 \textwidth]{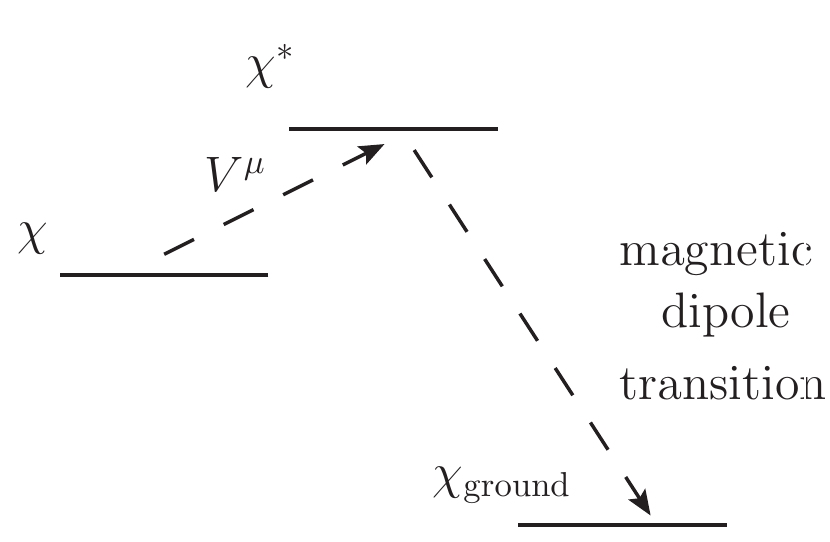}
\end{center}
\caption{The WIMP, $\chi$, may be part of a multiplet of states connected through a dark force, $V^\mu$. However, if the mass difference between the WIMP and the true ground state, $\chi_{\rm ground}$ is smaller than twice the electron mass then the WIMP is long-lived on cosmological time-scales. Its up-scattering against normal matter into the excited state, $\chi^*$, is then followed by a rapid decay to the ground state through the magnetic dipole transition. In this case, the energy threshold associated with the up-scattering is not the same as the energy associated with the decay down to the ground state. }
\label{fig:three-levels}
\end{figure}

Another interesting feature of such models is that the WIMP itself might be the long-lived metastable state if it is part of a multiplet of states in the dark sector, connected through a non-abelian force carrier. If the mass difference between the WIMP and the true ground state is less than twice the electron mass then this transition is highly suppressed~\cite{Finkbeiner:2009mi}. It was further shown in refs.~\cite{Finkbeiner:2009mi,Batell:2009vb} that such a state can have comparable abundance to the actual ground-state and is thus relevant for direct-detection experiments. It can scatter against normal matter into the excited state $\chi^*$ through the dark mediator as shown in Fig.~\ref{fig:three-levels}. If the excited state is coupled to the ground-state through a magnetic dipole it rapidly decays, emitting a characteristic x-ray. Importantly, the energy associated with the decay is not the same as the energy threshold associated with the scattering. Some concrete models realizing this setup are presented in the appendix. Rather than explore in detail all these different models, the purpose of this subsection is only to illustrate the broad range of models the search we propose in this paper is sensitive to. These models cogently illustrate that the scattering rate and decay rate may be uncorrelated. In addition, models with metastable states motivate searches for transition lines in the MeV range and not only in the usual 100 keV range. For simplicity, in the remainder of this paper we will concentrate on the possibility of the magnetic dipole moment as the only interaction for which both the scattering rate as well as the decay rate scale with the interaction strength, $\muX$. 

\section{Scattering rate in lead}
\label{sec:scattering_rate}

Normally one is interested in the recoil energy lost by the WIMP to the nucleus and the associated rate, Eq.~(\ref{eqn:dRdER_formula}). However, in our case we are actually interested in the recoil of the excited state of the WIMP itself after the collision and its final decay back to the ground state through the emission of a photon. This leads to some changes in the usual formalism, which we now address. Before we proceed, however, we note that since lead has a large charge $Z=82$, but a small magnetic dipole (only one isotope, $^{207}$Pb with $\mu({\rm ^{207}Hg})=0.58$ in units of nuclear magneton $\muN$) we can neglect the contribution to the scattering cross-section from the dipole-dipole interaction, $\sDD$. For the remainder of this work we will thus assume that the interaction is dominated by the dipole-charge scattering, Eq.~(\ref{eqn:sigmaDZ}). 

The average over the halo velocity distribution is slightly more involved than usual because we must express the nuclear recoil in terms of the excited state kinetic energy and only then average over the incoming WIMP velocity. Using conservation of energy in the lab frame we substitute $\ER = \frac{1}{2}\mX v^2 - \EX - \delta$ in Eq.~(\ref{eqn:sigmaDZ}) including the form-factor. The velocity average can then be expressed as an integral over the radial component which must be done numerically,
\onecolumngrid
\be
\nonumber
\left \langle g(v) \right \rangle &=& \frac{1}{\sqrt{\pi} \vobs\vbar}\frac{1}{N_{\rm esc}} 
\\
&\times& 
\left\{
%
% case 1 
%
\begin{array}{cc}
\int_{\vobs-\vesc}^{\vesc+\vobs}  dv~ v~g(v) \left( e^{-(v- \vobs)^2/\vbar^2} -e^{-\vesc^2/\vbar^2} \right)&   \vobs > \vesc,~ \vmin < \vesc-\vobs  \\
&
\\
%
% case 1 
%
\int_{\vmin}^{\vesc-\vobs}  dv~ v~g(v) \left( e^{-(v- \vobs)^2/\vbar^2} -e^{-(v+ \vobs)^2/\vbar^2} \right)&     \\
&
\\
+ \int_{\vesc-\vobs}^{\vesc+\vobs}  dv~ v~g(v) \left( e^{-(v- \vobs)^2/\vbar^2} -e^{-\vesc^2/\vbar^2} \right)&    \vobs < \vesc,~ \vmin < \vesc-\vobs  \\
&
\\
%
% case 3 
%
\int_{\vmin}^{\vesc+\vobs} dv~v~g(v) \left( e^{-(v- \vobs)^2/\vbar^2} -e^{-\vesc^2/\vbar^2} \right) & \vesc-\vobs < \vmin < \vesc+ \vobs \\
&
\\
%
% case 4
%
0  & \vobs+\vesc < \vmin    
\end{array}
\right.
\ee 
where, 
\be
N_{\rm esc} = {\rm Erf}(\vesc/\vbar)- \frac{2}{\sqrt{\pi}}\frac{\vesc}{\vbar}\exp(-\vesc^2/\vbar^2)
\ee
and the minimum velocity, $\vmin$, is given in terms of the excited WIMP kinetic energy, the nucleus' and WIMP's mass, and the inelasticity,
\be
\vmin ^2 = \frac{\mN^2+\mX^2}{(\mN-\mX)^2}~\frac{2 \EX}{  \mX }   \left( 1+\frac{\mN(\mN-\mX)}{\left(\mN^2+\mX^2\right)}\frac{\exE}{\EX}-2\frac{\mN \mX}{ \mN^2+\mX^2} \sqrt{1+\frac{\mN-\mX }{ \mN}\frac{\exE}{\EX}}\right)
\ee
\twocolumngrid

\begin{figure}
\begin{center}
\includegraphics[width=0.45 \textwidth]{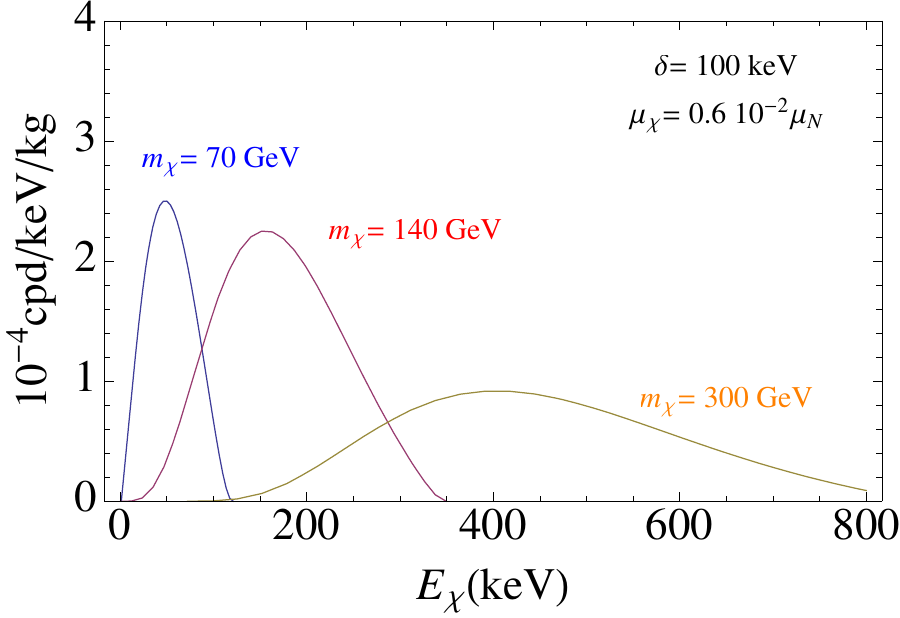}
\end{center}
\caption{The differential WIMP excitation rate, $dR/d\EX$ as a function of the excited WIMP energy $\EX$ for various values of the WIMP and a fixed value of the dipole moment and the inelasticity. These values correspond to the benchmarks presented in ref.~\cite{Chang:2010en} which results in good agreement with the DAMA signal. }
\label{fig:dRdEX}
\end{figure}

We note that the minimum velocity is such that $\vmin^2 \ge 2\exE/\mNX$ where $\mNX \equiv \mX\mN/(\mX+\mN)$ is the reduced mass of the WIMP-Nucleus system, which is expected from energy conservation in the COM frame. In addition, from the condition $\vmin < \vmax \equiv \vesc+\vobs$ we can derive the range of the excited state recoil energy spectrum,
\be
\nonumber
E_{\rm max/min} &=& \Bigg[ \frac{\mX^2+\mN^2}{\left(\mX+\mN \right)^2}  - \frac{\mN^2}{\left(\mX+\mN \right)^2} \frac{2\delta}{\mNX \vmax^2}  
\\
&\pm&  \frac{2\mNX^2}{\mX\mN} \sqrt{1-\frac{2\delta}{\mNX \vmax^2 }} \Bigg] \times  \frac{1}{2}\mX \vmax^2
\ee
The differential WIMP excitation rate is then  given by,
\be
\label{eqn:dRdEX}
\frac{dR}{d\EX} &=& {\rm N_{T}} \frac{\rhoX}{\mX}\left \langle v \frac{d\sigma}{d\EX}\right \rangle \\\nonumber&=& {\rm N_{T}} \frac{\rhoX}{\mX} \int_{\vmin} d^3 v f( \vec{v}+\vec{\vobs}) v \frac{d\sigma}{d\EX}.
\ee
As usual, the differential rate, $dR/d\EX$ modulates over the course of the year due to the motion of the Earth around the Sun~\cite{Drukier:1986tm, Chang:2011eb,Freese:2012xd}. However, daily modulations are much more significant as we discuss in the next section. In Fig.~\ref{fig:dRdEX} we show the unmodulated differential rate for some illustrative values of the mass, inelasticity, and magnetic dipole moment of the WIMP. We note that the rate scales with $\muX^2$. 

\section{Rate of decay of the excited WIMP in the detector}
\label{sec:decay_rate}

The collision rate in the lead shield is of course not what we are interested in at the end since it itself is not observed. After the WIMP collides in the shield and transition into the excited state, it travels some distance before decaying back to the ground state through the emission of a single photon of energy $\exE$. We are interested in the rate of such decays that occur inside the detector and that is what we set to calculate in this section. 

The calculation is straightforward in principle. The velocity of the outgoing excited WIMP in the lab frame, $\vec{u}_{\chi}$ is entirely determined by the incoming WIMP velocity, $\vec{v}$, and the angles of scattering in the centre-of-mass frame, $\{\cos\theta, \phi\}$. The incoming velocity is governed by the velocity distribution in the halo. The polar angle, $\cos\theta$, which is related to the energy of the WIMP in the lab, $\EX$, is distributed according to the differential cross-section, $d\sigma/d\EX$, whereas the azimuthal angle is uniformly distributed. The outgoing WIMP travels some distance before decaying at position $\vec{r}_{\rm dec}$ given by
\be
\vec{r}_{\rm dec} = \vec{r}_{\rm col} + \vec{u}_{\chi}t_{\rm dec}, 
\ee
where $\vec{r}_{\rm col}$ is the location of the collision in the lead shield, and $t_{\rm dec}$ is the decay time which is governed by the lifetime of the excited state~\cite{Chang:2010en},
\be
\label{eqn:lifetime}
\frac{1}{\tau} =  \frac{\muX^2 \exE^3}{\pi}
\ee
Convolving all the relevant distributions above and integrating over the independent variables (incoming velocity, outgoing direction, decay time, etc.), it is possible to obtain some analytic results for the rate of decays inside the detector\footnote{See for example refs.~\cite{Gondolo:2002np,Finkbeiner:2009ug,Lin:2010sb} for similar manipulations in the context of direct-detection of dark matter where the analytic expressions are rather simple.}. However, the corresponding expressions are complicated and the integrals must all be done numerically. Instead, in this section we present analytic results for the rate under the approximation of an isotropic emission from the collision site. This approximation, while entirely unjustified for any given collision site, gives the correct rate for a spherically symmetric shield and detector. More generally, it provides a good estimate for the total event rate even in the more realistic case of a shield which is not perfectly spherical, yet provides a $4\pi$ coverage of the detector. In the section that follows we present the result of a Monte-Carlo simulation that can be used for a general geometry, treats the kinematics exactly, and takes into account the anisotropic emission from the collision site. 

%%%%%%%%%%%%%%%%%%% subsection %%%%%%%%%%%%%%%%%%
%
%\subsection{Analytics - Overall Rate}

Under the assumption of a shield of uniform density and isotropic emission following the collision, the differential rate of collisions in a volume $V$ resulting in a WIMP of energy $\EX$ moving in the solid-angle $\Omega$ is
\be
\nonumber
\mathcal{R}_{\rm Pb} &\equiv& \frac{d^3R}{dVd\EX d\Omega} = \frac{\targetDensity}{4\pi}  \frac{dR}{d\EX} \\ &=& \frac{\targetDensity}{4\pi} \frac{\rhoX}{\mX}\left \langle v \frac{d\sigma}{d\EX}\right \rangle
\ee
where $\targetDensity$ is the number of targets available in a unit volume of lead. Thus, aside from an overall factor of $\targetDensity/4\pi{\rm N_T}$, we can simply use the results of the previous section, Eq.~(\ref{eqn:dRdEX}). The experimental observable is of course the rate of decays inside the detector, which is generally given by,
\be
\label{eqn:event_rate_formula}
R= \int d\EX \int d\Omega \int_{\rm shield} dV ~\mathcal{R}_{\rm Pb} \times P_{\rm d}\left(\EX,V,\Omega\right).
\ee 
Here, $P_{\rm d}\left(\EX,V,\Omega \right)$ is the probability that the excited state, which was produced at a volume $V$ in the shield with energy $\EX$ and angle $\Omega$ will arrive at the detector and decay inside it. Thus the problem neatly separates into two parts: the differential rate in lead, $\mathcal{R}_{\rm Pb}$, which is governed by the microscopic scattering rate; and $P_{\rm d}$, the probability of the excited state to decay inside the detector, which is determined by the geometry of the setup and the lifetime of the excited state. We note that this separation is only possible because we are ignoring the directional information associated with the motion of the lab with respect to the WIMPs in the halo. Including these effects would require to do the averaging over the halo velocities for every volume element in the shield since the result depends on the relative orientation of this element (with respect to the detector) compared with the WIMP wind. We address this issue in the next section using the Monte-Carlo simulation. 

The lifetime of the WIMP is given in Eq.~(\ref{eqn:lifetime}) above. In order to gain analytical control over the problem we model the detector as a sphere of radius $\rDet$ with the origin of the coordinate system in its center. The probability of the excited state to arrive at the detector and decay inside it is then
\be
\label{eqn:decay_probability}
P_{\rm decay} = \exp\left(-\frac{t_{\rm enter}}{\tau}\right) - \exp\left(-\frac{t_{\rm exit}}{\tau}\right)
\ee
where $t_{\rm enter}$ ($t_{\rm exit}$) is the time the excited state entered (exited) the detector after the initial collision. If the initial collision occurred a distance $r$ away from the centre of the detector and the excited state was moving at an angle $\theta$ with respect to this direction, then the times are given by,
\be
\label{eqn:time_in_detector}
t_{\rm exit/enter} = \frac{r \cos\theta \pm \sqrt{\rho^2 - r^2 \sin^2\theta}}{v_\chi}.
\ee
The velocity of the excited state is $v_\chi = \sqrt{2 \EX\mXp}$ and we note that the maximum angle that would reach the detector is given by $\sin\theta_{\rm max}  = \rho/r$. 

\begin{figure}
\begin{center}
\includegraphics[width=0.45 \textwidth]{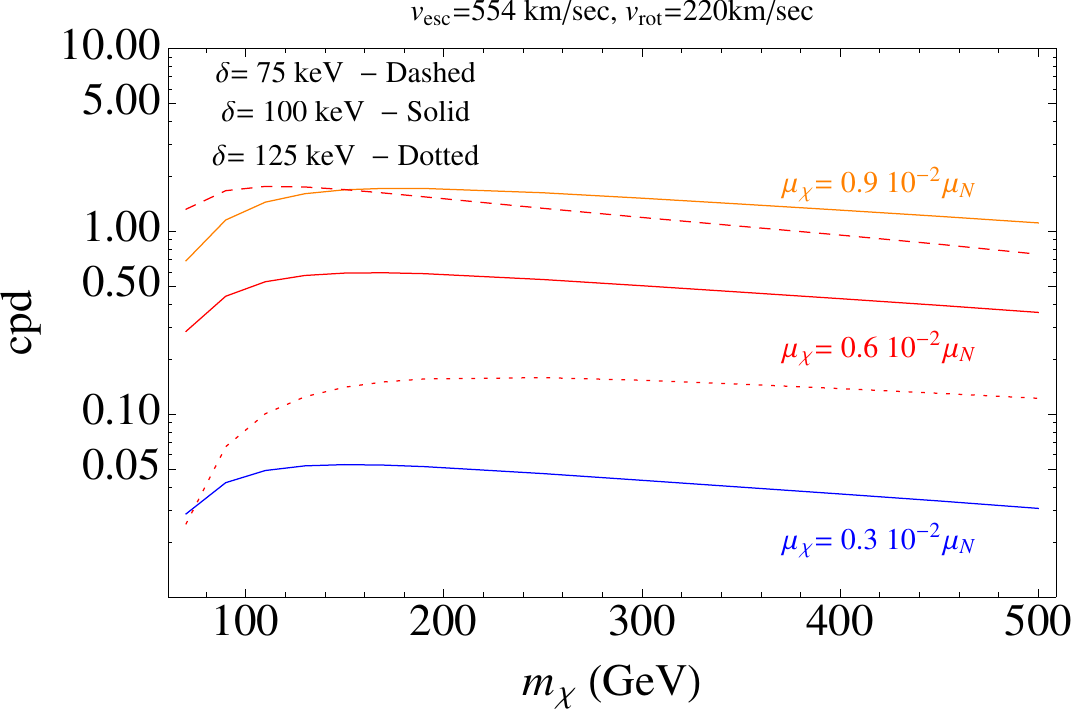}
\end{center}
\caption{The rate (counts per day) as a function of the WIMP mass for several values of the magnetic dipole moment. The spherical shield and detector were taken to have approximately the same volume as those of the CUORE-0 and CUORICINO experiments~\cite{Pavan:2012cta}: the shield was a spherical shell with outer radius of 61 cm and an inner radius of 35 cm; the spherical detector has a radius of 12 cm. The inner most shield in the CUORE-0 setup was neglected.}
\label{fig:rate_in_cuore0}
\end{figure}

\begin{figure}
\begin{center}
\includegraphics[width=0.45 \textwidth]{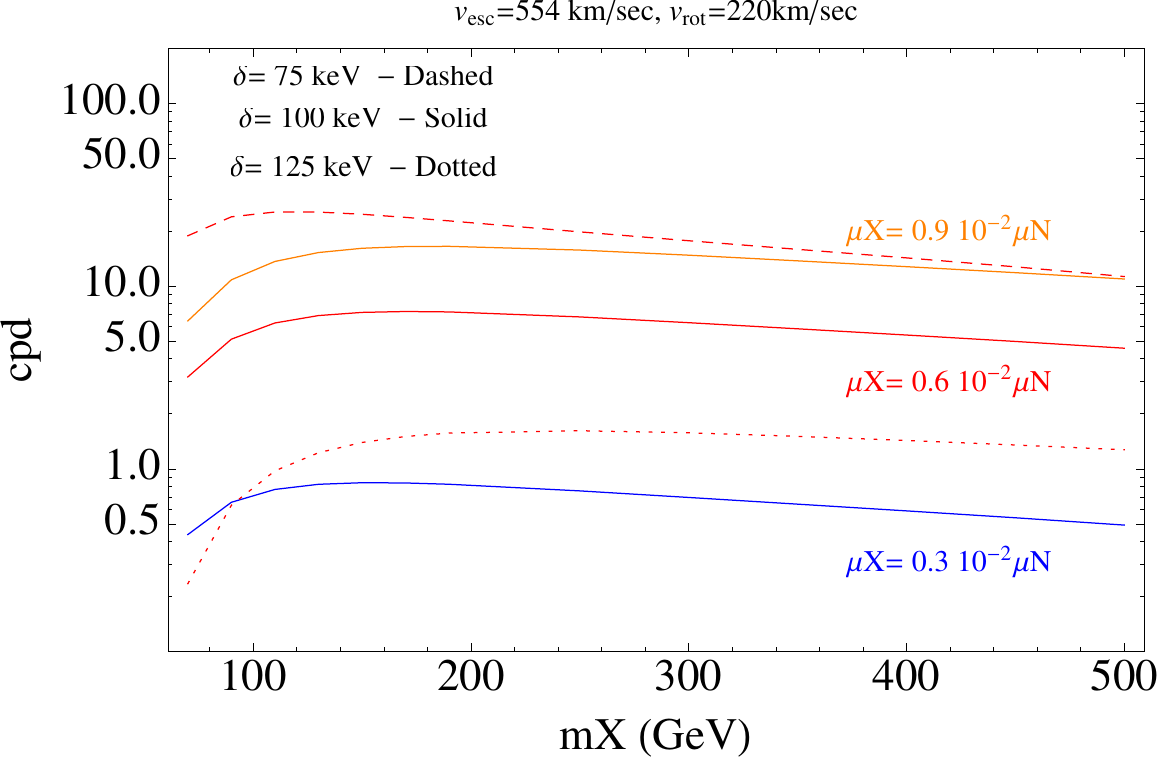}
\end{center}
\caption{The rate (counts per day) as a function of the WIMP mass for several values of the magnetic dipole moment. The spherical shield and detector were taken to have the same volume as those of the CUORE experiment~\cite{Arnaboldi:2002du}: the outer shield is modeled as a spherical shell with an outer radius of 118 cm and an inner radius of 93 cm; the inner shield has an outer radius of 35 cm and an inner radius of 31 cm; the spherical detector has a radius of 31 cm.}
\label{fig:rate_in_cuore}
\end{figure}

Using Eqs.~(\ref{eqn:decay_probability}) and (\ref{eqn:time_in_detector}) we can explicitly evaluate the event rate, Eq.~(\ref{eqn:event_rate_formula}) for a given spherical geometry. As an example, in Fig.~\ref{fig:rate_in_cuore0} we show the event rate in a spherical detector/shield configuration with the same volumes as those of the CUORICINO and CUORE-0 setups~\cite{Pavan:2012cta}. The actual experimental configuration is of course not spherical, an issue we address in the next section, but the figure gives a quick and reasonable estimate for the expected event rate for different WIMP model parameters. Similarly, in Fig.~\ref{fig:rate_in_cuore} we show the event rate in a spherical geometry based on the CUORE setup~\cite{Arnaboldi:2002du}. The cubical CUORE setup is indeed somewhat closer to being spherical and we confirmed that these analytic results are in good agreement with the simulation results presented in the next section. Using the more precise cubical geometry of the detector and shield leads to about a 30\% reduction in the rate. 

As can be expected, the event rate in the "CUORE"-like setup, Fig.~\ref{fig:rate_in_cuore},  is about an order of magnitude larger than that in the "CUORICINO"-like setup, Fig.~\ref{fig:rate_in_cuore0}. However, this does not necessarily imply a greater sensitivity since the total background inside the detector also scales with its size. A precise evaluation of the sensitivity would require a detailed knowledge of the energy resolution and the background level at each energy bin. Independently of these considerations, the sensitivity can be further increased by considering the temporal shape of the signal.

%%%%%%%%%%%%%%%%%% section %%%%%%%%%%%%%%%%%%
\section{Monte-Carlo Simulation - diurnal modulations}
\label{sec:monte_carlo}

\begin{figure}
\begin{center}
\includegraphics[width=0.45 \textwidth]{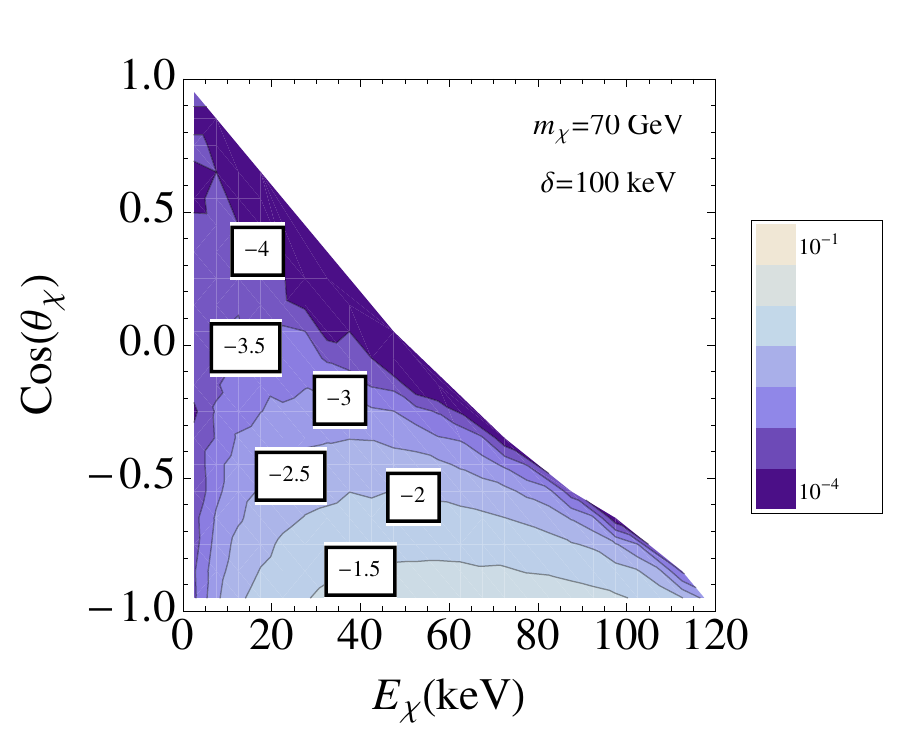}
\includegraphics[width=0.45 \textwidth]{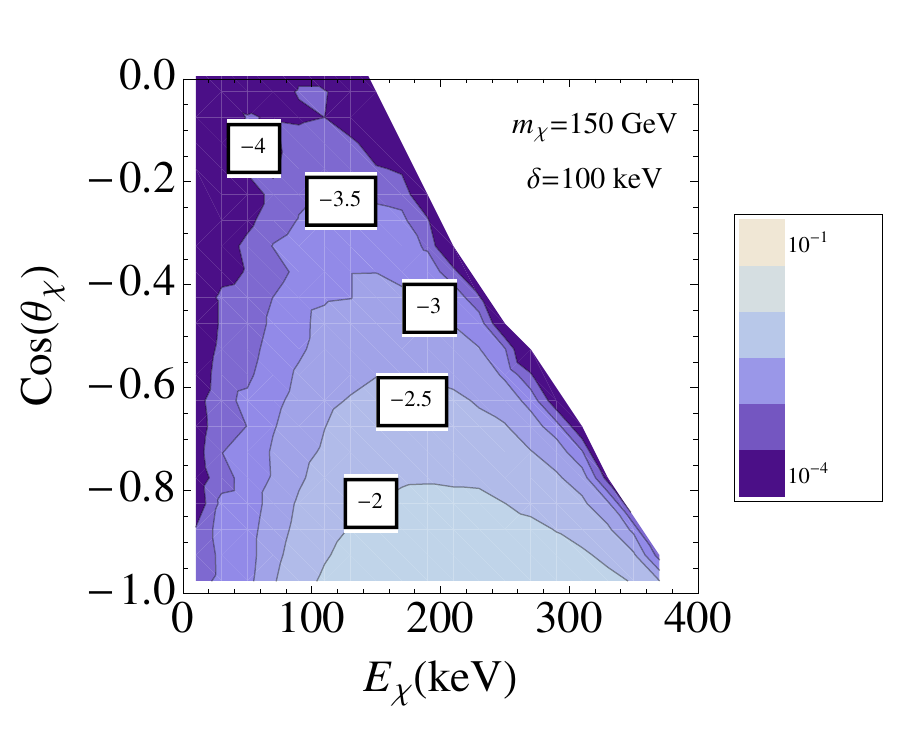}
\includegraphics[width=0.45 \textwidth]{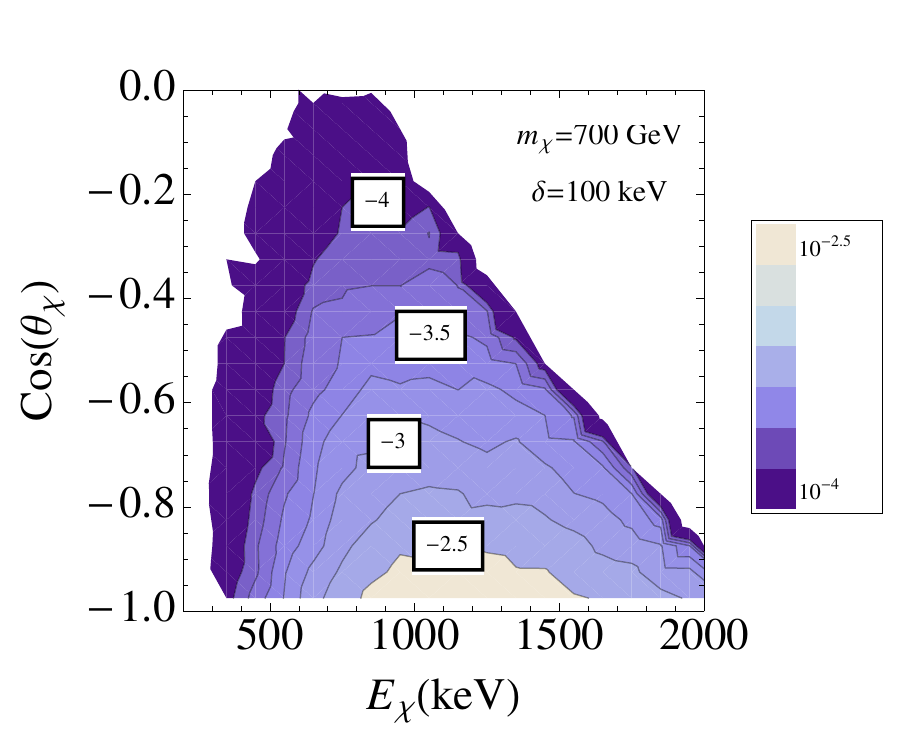}
\end{center}
\caption{The normalized differential rate with respect to both the energy and direction of the excited WIMP, $d^2R/d\EX d\cos\thetaX$ for various choices of the mass and inelasticity.}
\label{fig:dRdEXdCosX}
\end{figure}

The above results provide a good estimate for the number of events expected and demonstrate the great potential of existing experimental efforts to look for such phenomenon over a wide range of the relevant parameter space. However, the angular distribution of the WIMP leaving the collision site is in fact not isotropic and carries with it useful information which can be utilized to achieve greater experimental sensitivity. The angle, $\thetaX$, of the outgoing velocity, $\vec{u}_\chi$, is defined with respect to the velocity of the observer in the halo, 
\be
\hat{u}_\chi \cdot \hat{v}_{\rm o} = \cos\thetaX
\ee
Intuitively, we expect that when the WIMP is much heavier than the nucleus it should not deviate much from its original direction. The average incoming direction is opposite to the direction of the observer in the halo, $\langle \vec{v} \rangle = -\vec{\vobs}$. Therefore we expect $\thetaX$ to peak around $\pi$, as indeed it does: in Fig.~\ref{fig:dRdEXdCosX} we show the results of a simulation of the velocity averaged differential rate with respect to both the energy and direction of the excited WIMP, $d^2R/d\EX d\cos\thetaX$, for various choices of the mass and inelasticity. The distribution is sharply peaked towards $\cos\thetaX = -1$ unless the WIMP is considerably lighter than the lead nucleus. 

The direction of the outgoing excited state is therefore strongly correlated with the direction of the WIMP wind. This results in a strong correlation between the geometry of the experimental setup and the event rate as a function of time. As the lab rotates diurnally with respect to the WIMP wind, different parts of the lead shield come between the WIMP wind and the detector. If the shield geometry is highly anisotropic then the event rate undergoes large diurnal modulations. These modulations have a well-defined characteristic shape that depends on the known geometry of the experimental apparatus. 

As is well known, but we reiterate, because the period of this modulation is that of a sidereal day (23h 56 min), it can be definitively separated from effects that could arise with a solar day (24h) period.

%%%%
%%%%

Every sensitive $0\nu2\beta$ decay experiment may detect previously unidentified/unassigned mono-energetic lines (see, {\em e.g.} Refs. 
\cite{Gando:2012zm,Alessandria:2012zp}
for some recent examples). Since the value of $\delta$ is a completely free parameter for us, such a background line within a single detector 
can easily be confused with a signal. One way to counter this is to compare different detectors, where a true signal would appear at $E=\delta$ and be independent on the levels of background. However, it is clear that even within a single detector one can use diurnal modulations to separate presumably constant or very weakly time-dependent backgrounds from the WIMP de-excitation signal. Thus, the diurnal modulation present a formidable tool in enhancing the sensitivity of the search.  

%%%%
%%%%

\begin{figure}[t]
\begin{center}
\includegraphics[width=0.45 \textwidth]{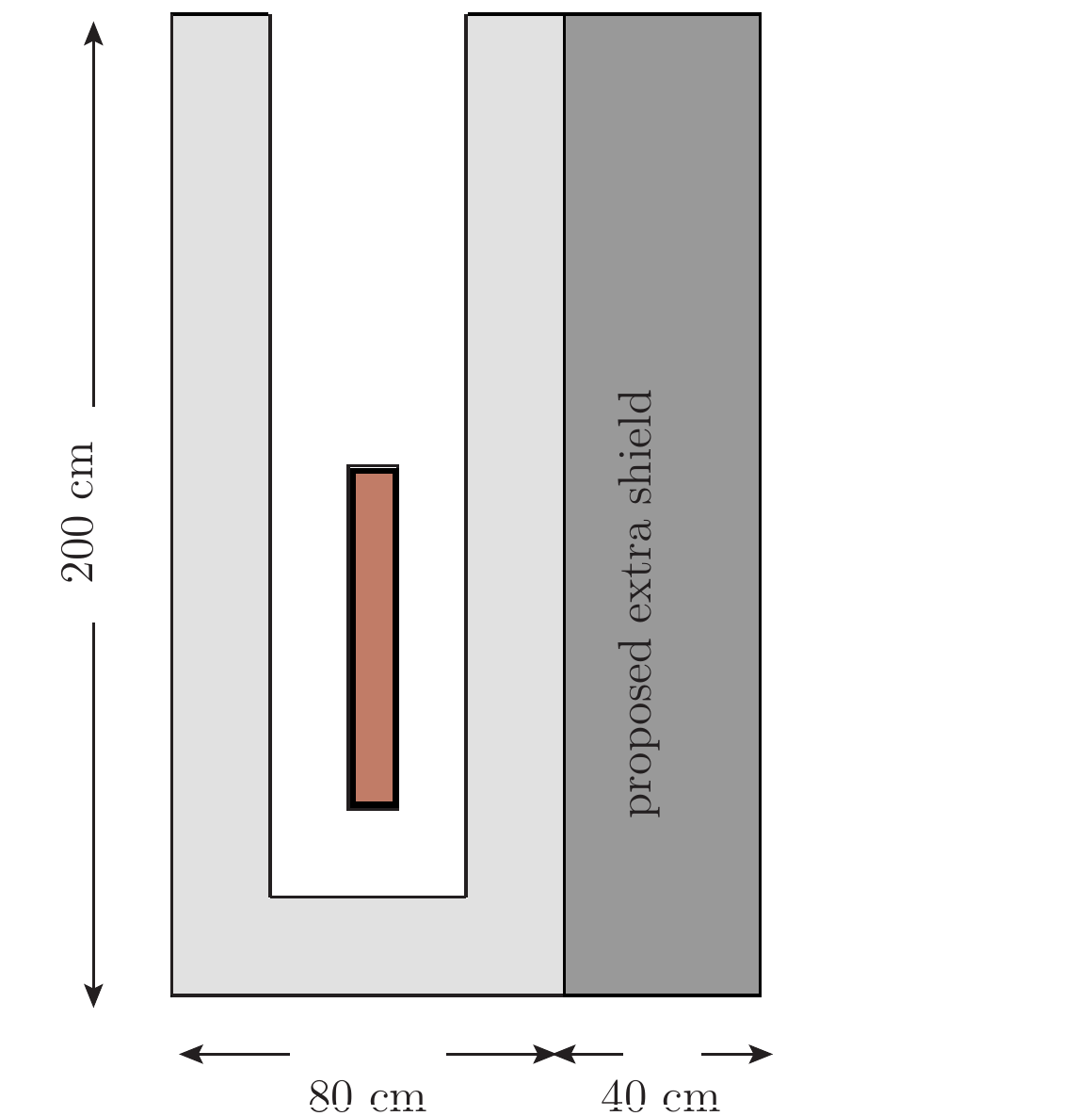}
\end{center}
\caption{Our simulated experimental setup based schematically on the CUORICINO/CUORE-0 setup~\cite{Arnaboldi:2008ds}. The detector is a $10 \times 10 \times 70~{\rm cm}^3$ positioned 10 cm above the bottom shield. It is surrounded by an inner lateral shield of 1.2 cm thick layer of Roman lead (not depicted). We neglect the small additional upper shield present directly above the detector as it contributes very little to the total rate. The outer shield is 20 cm thick and modeled as depicted. On the right we depict our proposed additional lead shield, $40 \times 80 \times 200~{\rm cm}^3$ facing north, which as discussed in the text, greatly increases the overall rate and diurnal modulations.   Diagram is not drawn to scale.  }
\label{fig:CUORICINO_schematic}
\end{figure}

\begin{figure}[t]
\begin{center}
\includegraphics[width=0.45 \textwidth]{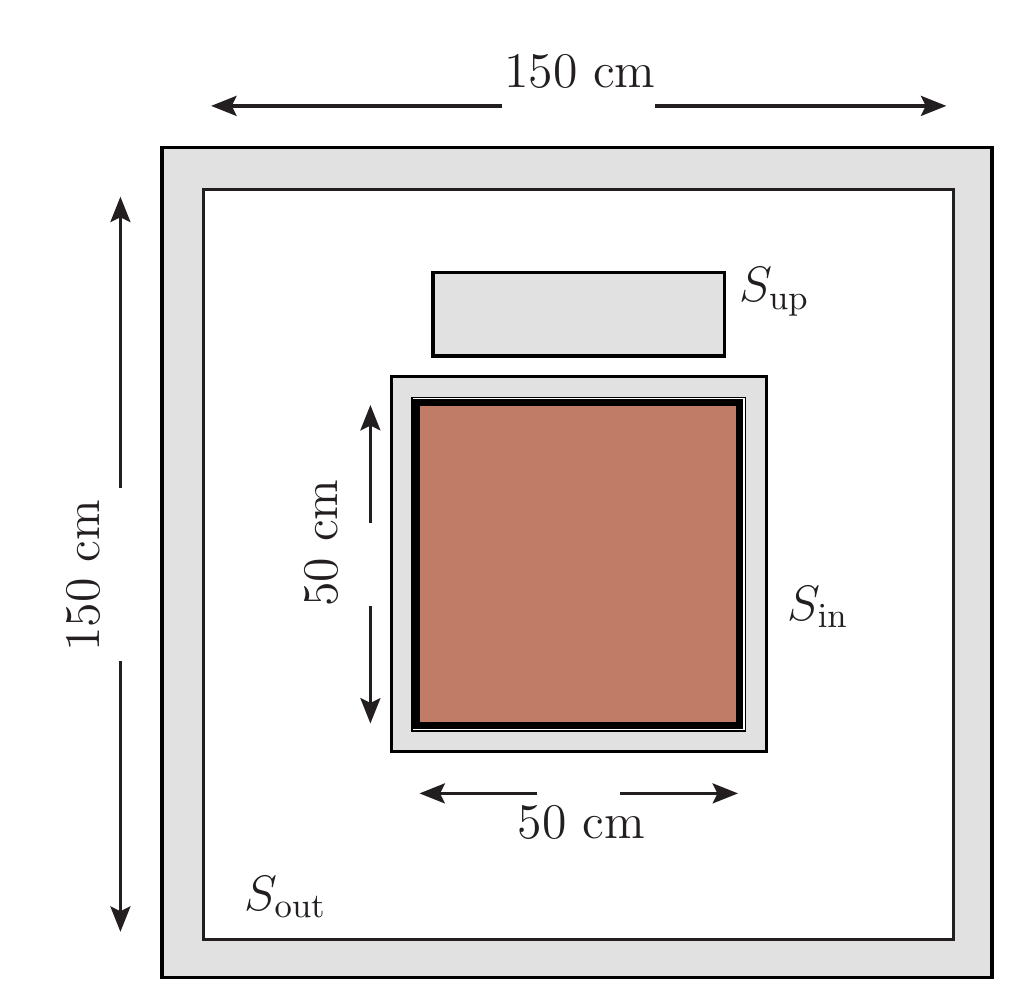}
\end{center}
\caption{Schematics of our simulated experimental setup based on the CUORE setup shown in Fig.~9 of ref.~\cite{Arnaboldi:2002du}. The detector is a $50 \times 50 \times 50~{\rm cm}^3$ cube surrounded by the inner shield, $S_{\rm in}$ - a 3 cm thick layer of Roman lead. We model the outer shield, $S_{\rm out}$, as a cube with an inner length of $1.5$ meters and a thickness of 20 cm with the detector in its center. This is not precise, but sufficient for the purpose of this paper. The additional upper shield present directly on top of the upper face of the detector has an area of $40\times 40~{\rm cm}^2$ and is 17 cm thick. This asymmetric configuration enhances the diurnal modulations as discussed in the text. Diagram is not drawn to scale.  }
\label{fig:CUORE_schematic}
\end{figure}

In order to investigate this effect, we developed a simple Monte-Carlo simulation based on two experimental setups. The first, shown in Fig.~\ref{fig:CUORICINO_schematic}, is schematically based on the past CUORICINO~\cite{Arnaboldi:2008ds} experiment and the present CUORE-0~\cite{Rusconi:2012dfa} setup. The second, shown in Fig.~\ref{fig:CUORE_schematic}, is based on the planned CUORE experimental setup as presented in Fig.~9 of ref.~\cite{Arnaboldi:2002du}. In the following we discuss each case separately and present the expected rates and temporal modulations for each setup. However, we emphasize that while the CUORE experiments are particularly well-suited for our purpose, other experiments based on different detection techniques and materials can similarly be used to search for this phenomenon. Indeed, our simulation relied only on the geometry of the lead shield and detector, but not on the specific detection technique used by the experiment (except to assume that x-rays can be observed with sufficient fidelity). 

\begin{figure}[t]
\begin{center}
\includegraphics[width=0.5 \textwidth]{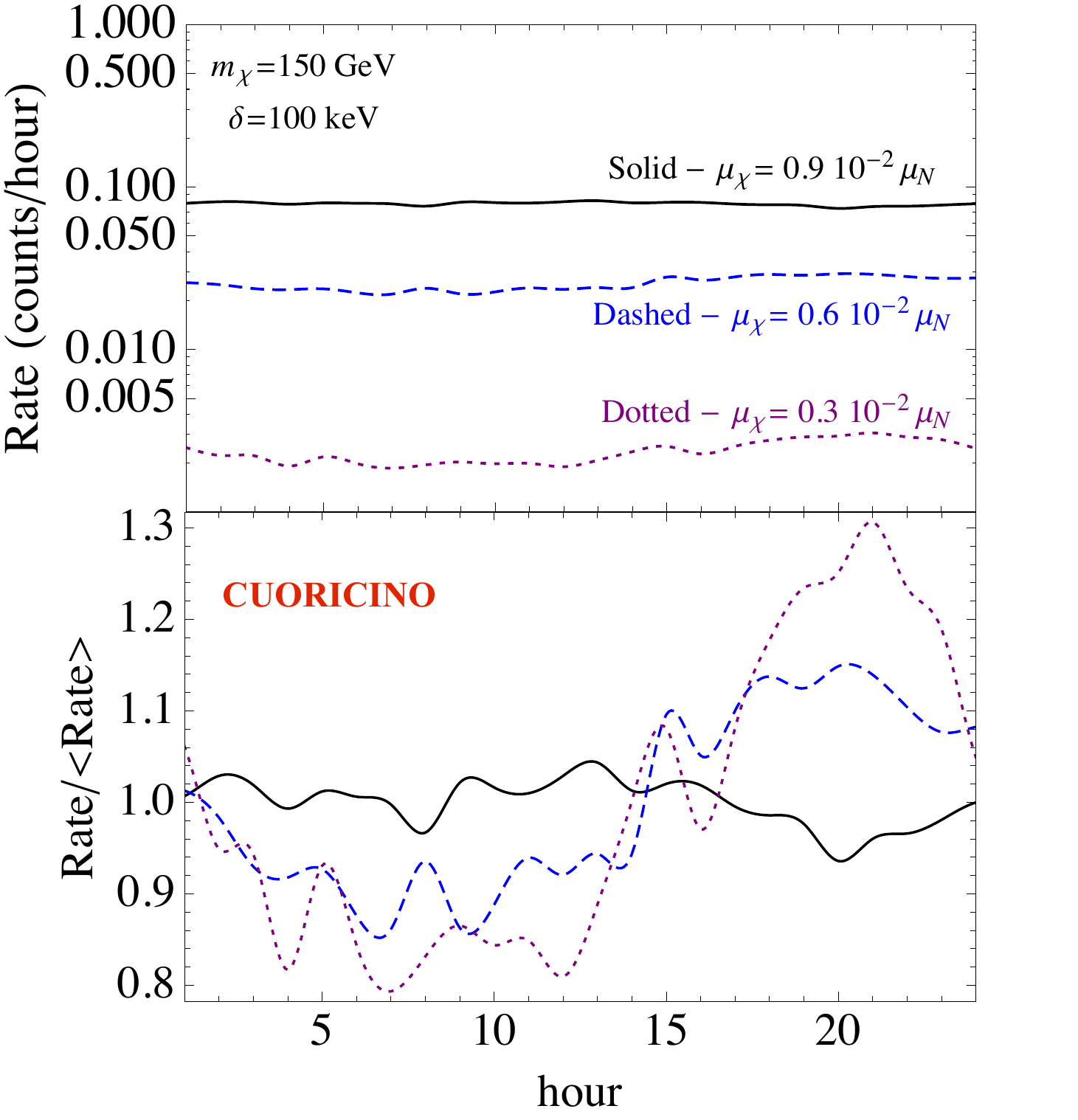}
\end{center}
\caption{In the upper pane we plot the number of counts per hour in CUORICINO/CUORE-0, Fig.~\ref{fig:CUORICINO_schematic}, against the GAST hour of the day for various choices of the dipole moment. In the lower pane, we show the expected diurnal modulations by plotting the hourly rate divided by the average rate for the same parameters. The lifetimes corresponding to $\muX = (0.3, 0.6, 0.9)\times 10^{-2}~\mu_{\rm N}$ are $(1.0, 2.2, 9.0)~{\rm \mu s}$, respectively.  }
\label{fig:fraction_vs_muX_CUORICINO}
\end{figure}

\begin{figure}[t]
\begin{center}
\includegraphics[width=0.45 \textwidth]{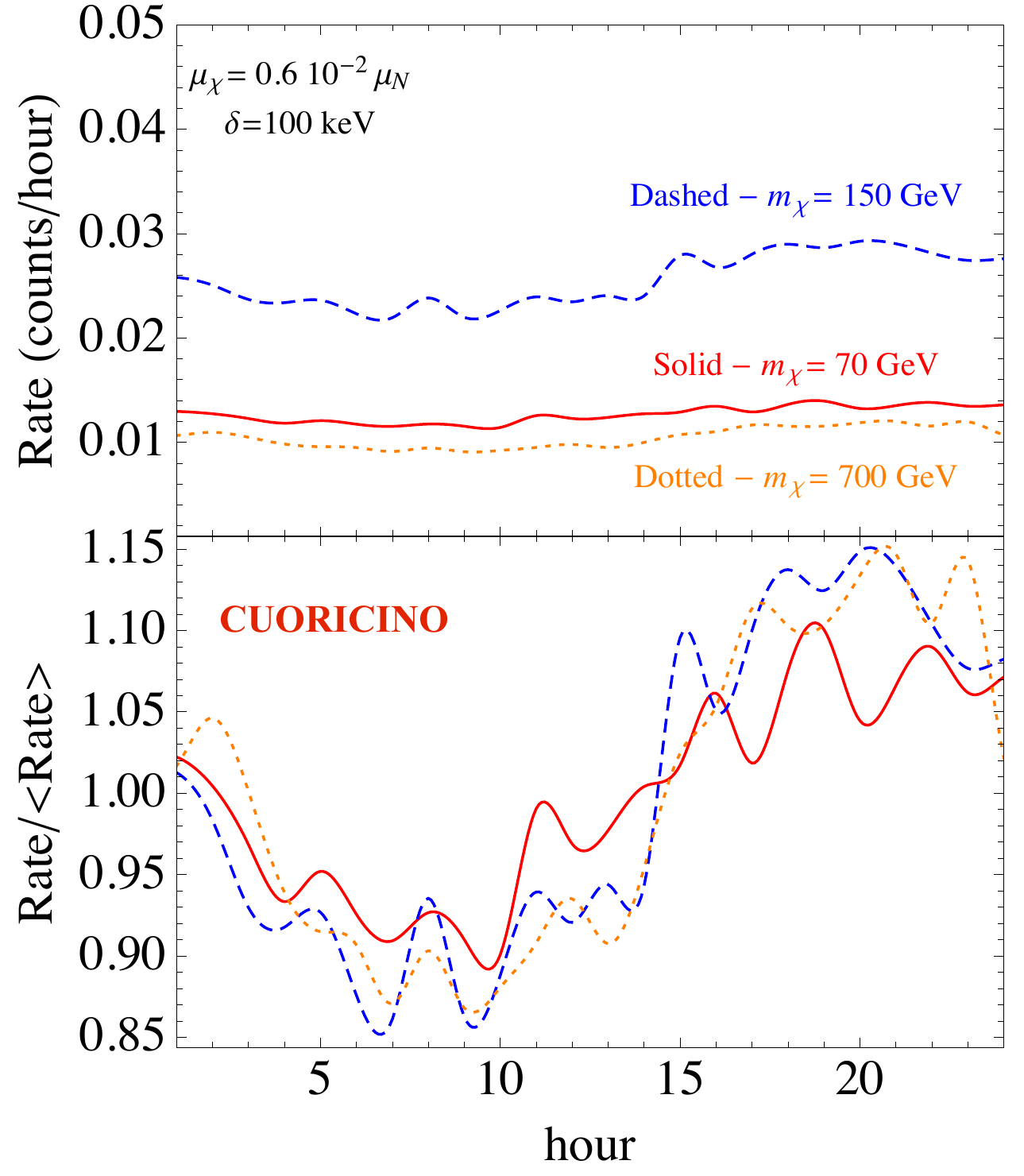}
\end{center}
\caption{Same as Fig.~\ref{fig:fraction_vs_muX_CUORICINO}, but for various choices of the WIMP's mass.}
\label{fig:fraction_vs_mX_CUORICINO}
\end{figure}

We begin with the first setup based on the CUORICINO/CUORE-0 experiments shown in Fig.~\ref{fig:CUORICINO_schematic}. This setup has a strong asymmetry between the horizontal and vertical directions and one can expect strong diurnal modulations. However, for excited WIMPs with a very short lifetime only the geometry of the shield directly next to the detector matters and in that case the asymmetry is reduced since the different directions are not too dissimilar. This can be seen in Fig.~\ref{fig:fraction_vs_muX_CUORICINO} where we show the total rate expected as well as the fractional modulation against the hour of the day, for different values of the magnetic moment of the WIMP, $\muX$ (which in turn controls the lifetime through Eq.~(\ref{eqn:lifetime})). The hour is defined with respect to Greenwhich apparent sidereal time (GAST), which is related to the local apparent sidereal time in the lab through $t_{\rm lab} = t_{\rm GAST} + l_{\rm lab}/15$ where $l_{\rm lab} = 13.7^o$ is the longitude of Gran-Sasso lab (more details on the different time frames can be found in ref.~\cite{Bozorgnia:2011tk}).  As expected, the overall rate is lower for lower values of the magnetic moment since fewer collisions occur in the shield. On the other hand, the diurnal modulations are large, $\sim 30\%$, for smaller values of the magnetic moment as the lifetime becomes longer.  In Fig.~\ref{fig:fraction_vs_mX_CUORICINO} we again show the total expected hourly rate as well as the modulations, but for different choice of the WIMP's mass and a fixed value of the magnetic moment. As can be expected, while the overall rate is somewhat sensitive to the mass the modulations are not.

\begin{figure}[t]
\begin{center}
\includegraphics[width=0.55 \textwidth]{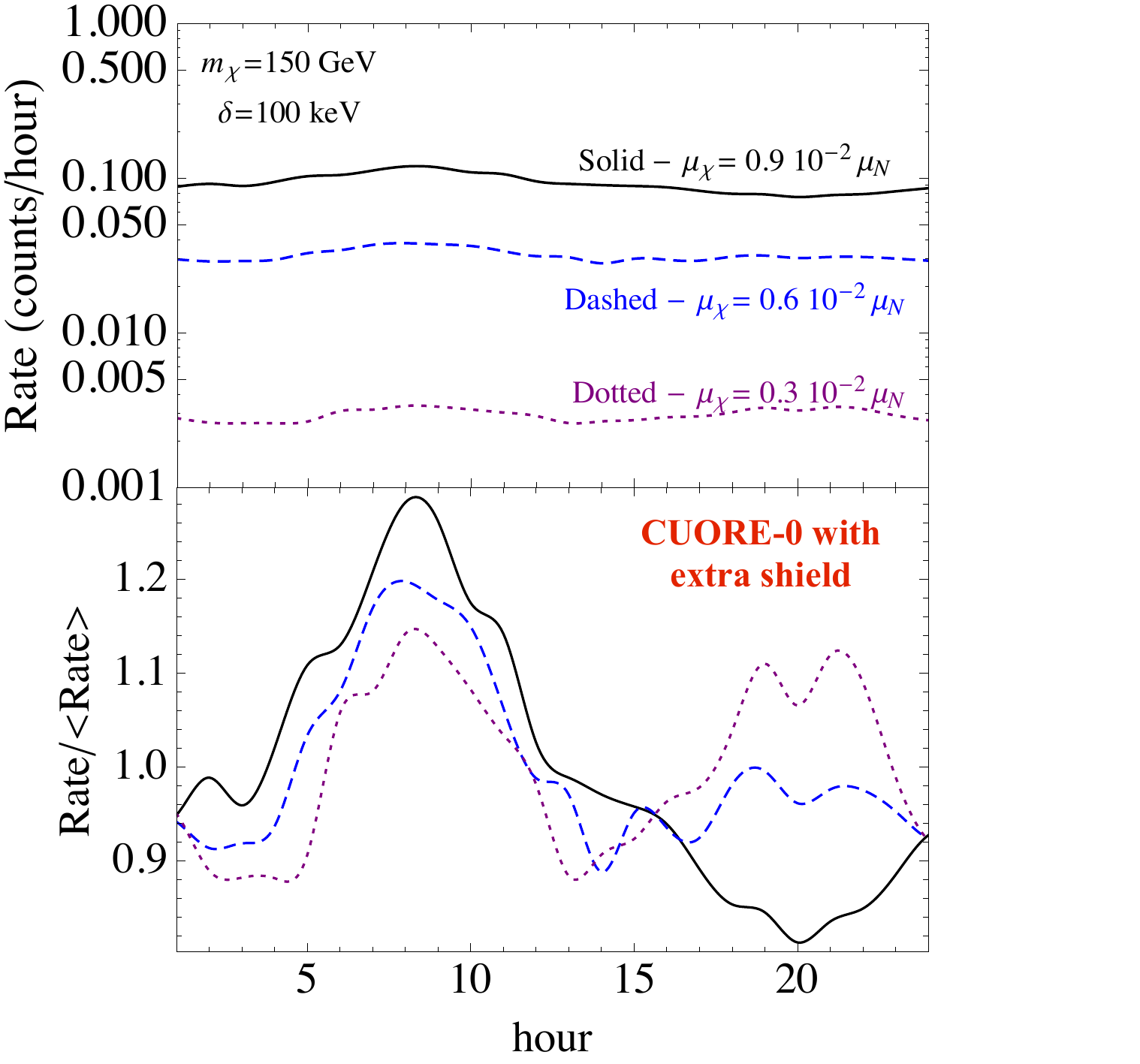}
\end{center}
\caption{In the upper pane we plot the number of counts per hour in CUORICINO/CUORE-0 with the additional northern shield, Fig.~\ref{fig:CUORICINO_schematic}, against the GAST hour of the day for various choices of the dipole moment. In the lower pane, we show the expected diurnal modulations by plotting the hourly rate divided by the average rate for the same parameters.}
\label{fig:fraction_vs_muX_CUORE0}
\end{figure}

Both the overall rate as well as the modulation fraction can be enhanced in the present CUORE-0 experiment with an additional external shield. Ideally, the additional shield should be made as large as possible, placed as close as possible to the detector, and designed to complement the anisotropies of the existing geometry to enhance both the rate and modulation fraction for long as well as short lifetimes. While such a shield can certainly be designed for any given experiment, any such extra shielding is likely to be constrained by the existing setup (and purpose) of the experiment. Therefore, in Fig.~\ref{fig:CUORICINO_schematic} we propose a simple addition in the form of a large external shield facing north and placed directly adjacent to the existing external shield. In Fig.~\ref{fig:fraction_vs_muX_CUORE0} we plot the total rate of excited state decays in the detector as well as the diurnal modulation fraction. The total rate is increased by about 30\% as compared to the regular setup. More importantly, the modulations fraction are now large even in the case of a short lifetime of the excited WIMP. On the other hand, since the extra northern facing shield causes an increased rate in the early hours, it destructively interferes with the modulations expected in the case of long lifetimes. Thus, a possibly better addition would be an extra shield positioned directly underneath the detector, which would nicely complement the existing anisotropy of the apparatus.

\begin{figure}[t]
\begin{center}
\includegraphics[width=0.45 \textwidth]{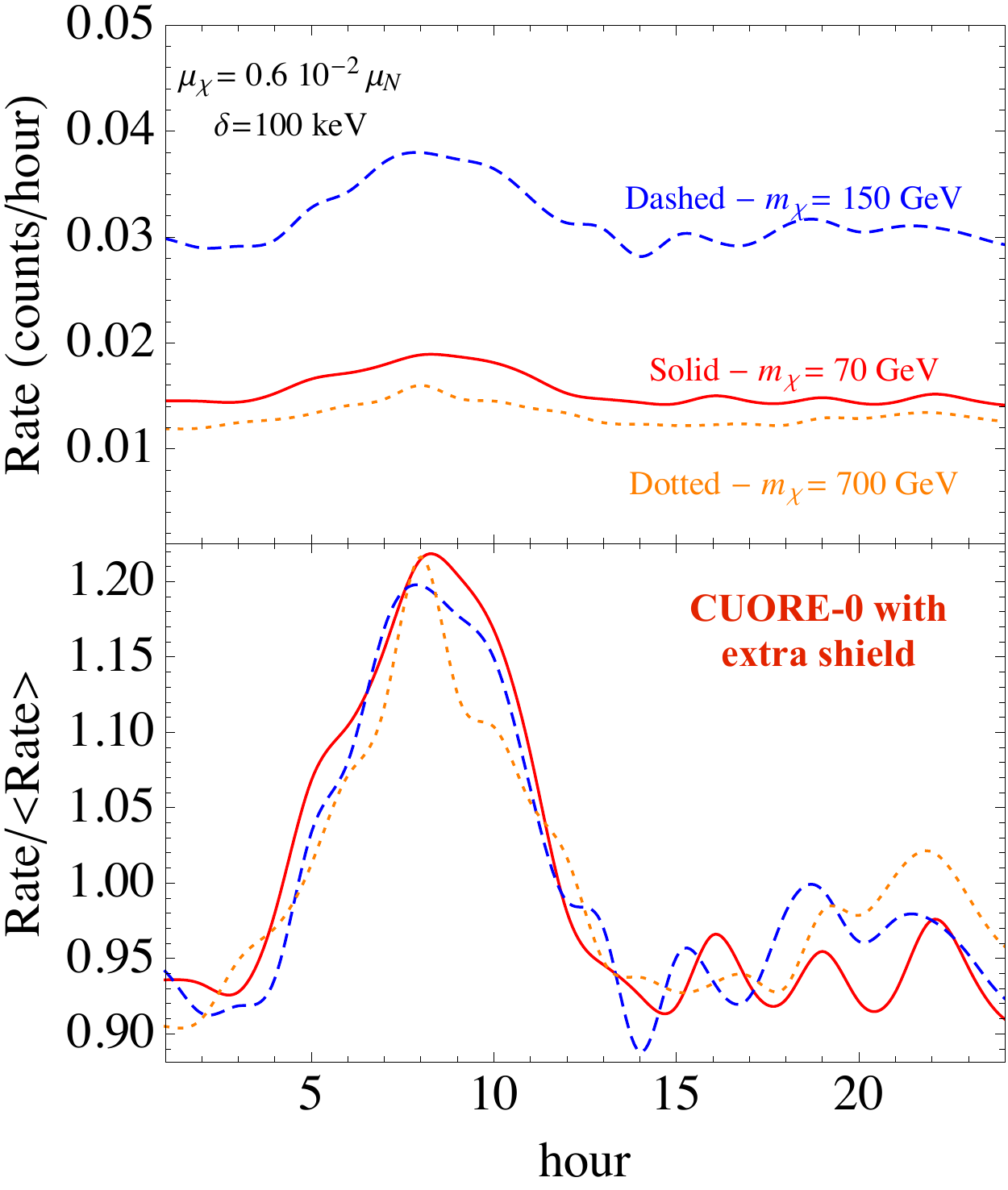}
\end{center}
\caption{Same as Fig.~\ref{fig:fraction_vs_muX_CUORE0}, but for various choices of the WIMP's mass.}
\label{fig:fraction_vs_mX_CUORE0}
\end{figure}

\begin{figure}[t]
\begin{center}
\includegraphics[width=0.45 \textwidth]{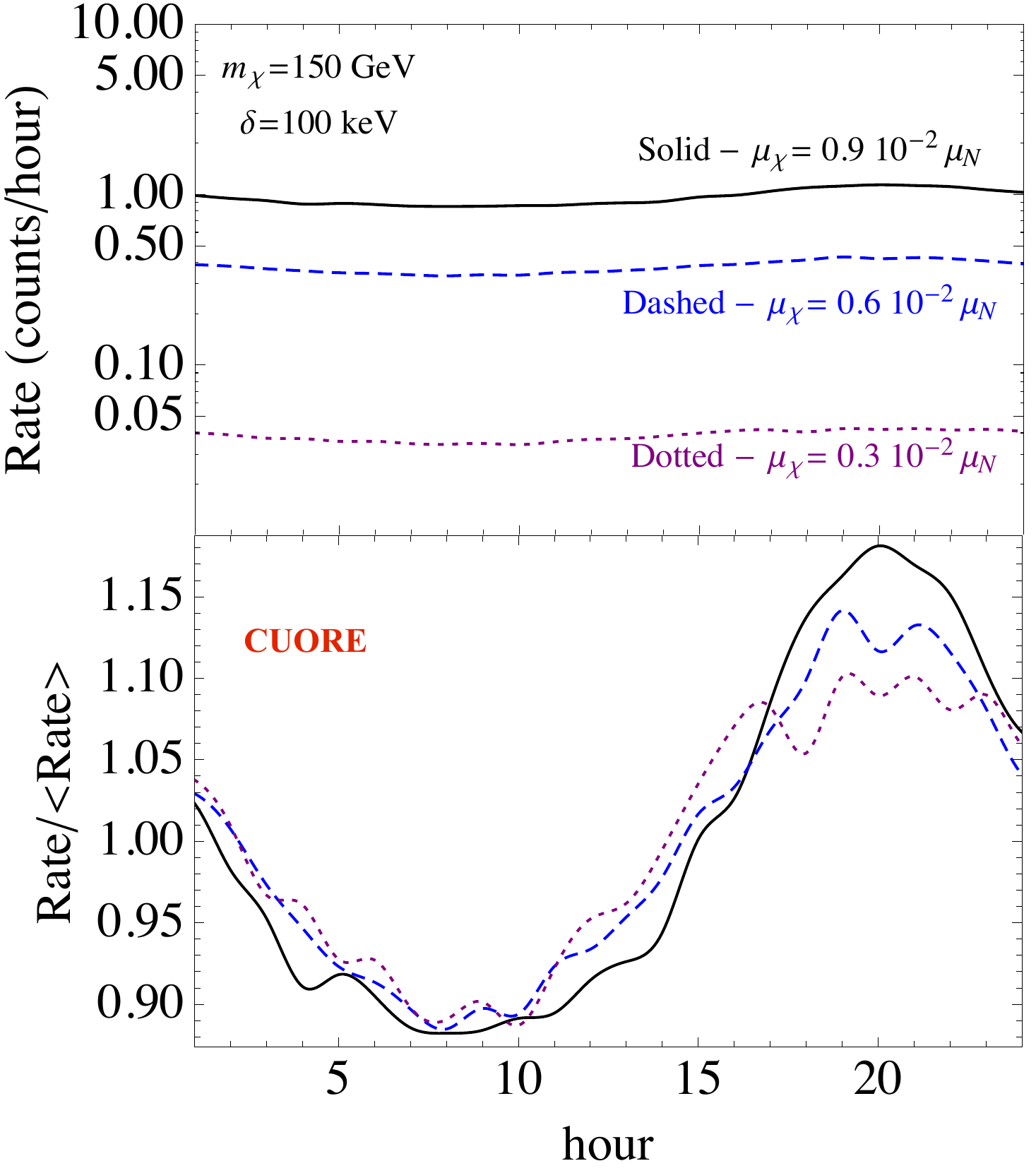}
\end{center}
\caption{In the upper pane we plot the number of counts per hour in CUORE, Fig.~\ref{fig:CUORE_schematic}, against the hour of the day for various choices of the dipole moment. In the lower pane, we show the expected diurnal modulations by plotting the hourly rate divided by the average rate for the same parameters.}
\label{fig:fraction_vs_muX_CUORE}
\end{figure}

\begin{figure}[t]
\begin{center}
\includegraphics[width=0.44 \textwidth]{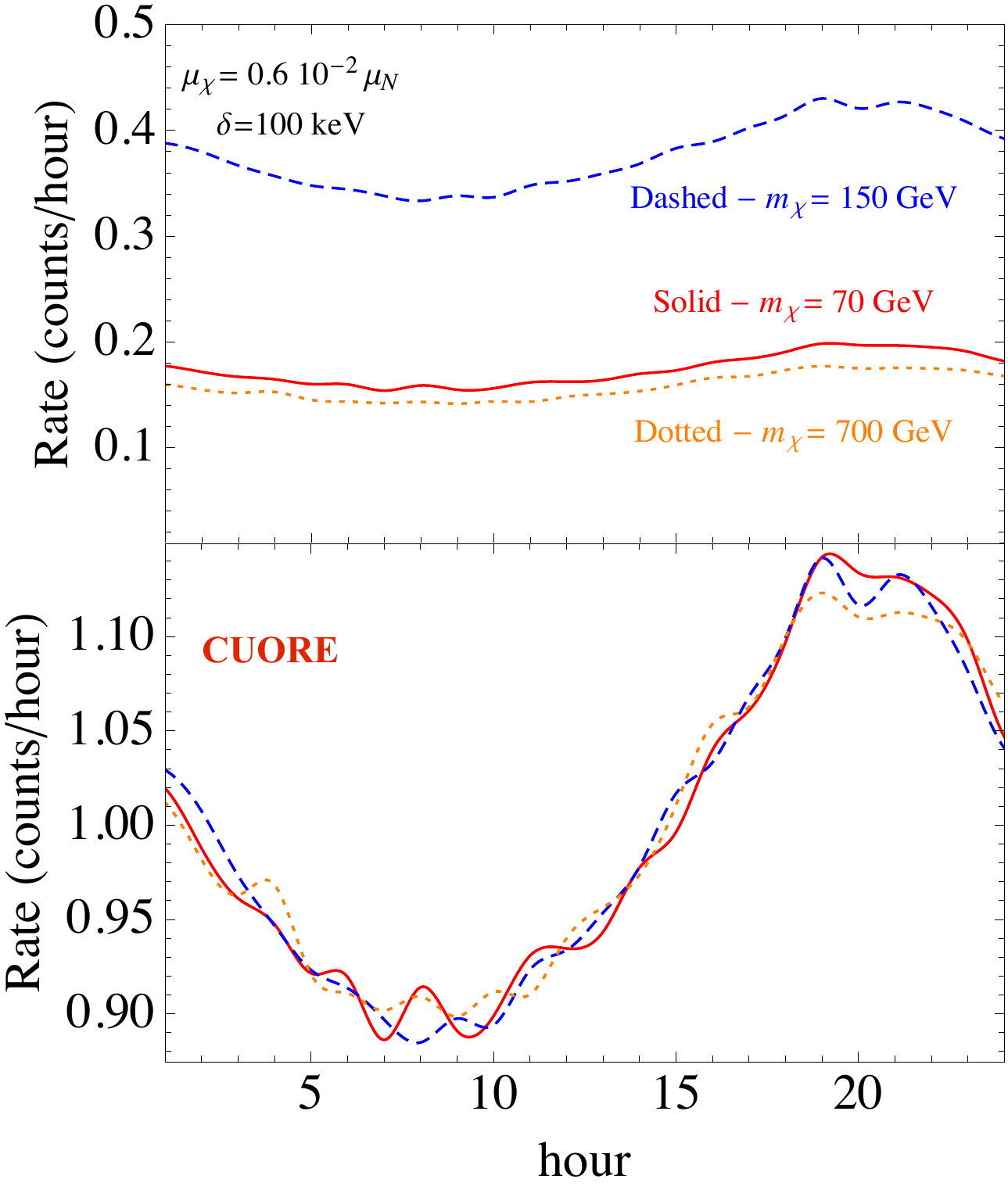}
\end{center}
\caption{Same as Fig.~\ref{fig:fraction_vs_muX_CUORE}, but for various choices of the WIMP's mass.}
\label{fig:fraction_vs_mX_CUORE}
\end{figure}

Next we consider the second setup based on the planned CUORE experiment, Fig.~\ref{fig:CUORE_schematic}. The cubical detector is surrounded by an inner lateral shield and a large external shield. This configuration is fairly isotropic and does not lead to large diurnal modulations. However, an additional shield is placed on the top face of the detector and results in sizable fractional modulations of $\sim 15\%$. This is shown in Fig.~\ref{fig:fraction_vs_muX_CUORE} where we plot the expected hourly rate of excited WIMPs decaying inside the detector alongside the fractional modulations. We note that the overall rate is increased by about an order of magnitude as compared with the CUORICINO/CUORE-0 experiments, as was anticipated in the previous section. In this case the modulations are in fact somewhat diminished for WIMPs with a longer lifetime since more of the, approximately isotropic, outer shield contributes to the signal. In Fig.~\ref{fig:fraction_vs_mX_CUORE}  we compare the total rate and modulations for various choices of the WIMP's mass. As before, while the total rate is sensitive to the mass, the modulation is not. 

\section{Conclusions}
\label{sec:conclusions}

In this paper we proposed a new search for WIMPs with a nearby excited state that can be done with existing experiments. The inelastic collision of a WIMP with the lead nuclei inside the ubiquitous lead shield present in experiments results in an excited WIMP state which subsequently decays via a single photon emission in the detector. Such decays would be seen as monochromatic x-ray line corresponding to the excited state energy. While the precise sensitivity depends on the detector resolution and background level at every energy bin, we showed that it can be greatly enhanced by utilizing the expected sidereal daily modulations of the signal. The direction of the excited state leaving the collision site is strongly correlated with the direction of the laboratory with respect to the WIMP wind. Thus, the daily rotation of the Earth together with the anisotropy of the experimental apparatus (shield and detector) leads to a strong diurnal modulations with a characteristic signature. 

We emphasize that while the interpretation of such a modulation (or the constraints arising from its absence) can be very model-dependent, the discovery of such \textsl{sidereal} daily modulations would definitively indicate the existence of dark matter. Given the relatively low cost to implement such a search, it seems a natural component of any experiment with high $\gamma$-energy resolution going forward. In fact, considering how little we know about the nature of dark matter, it seems prudent to look for \textsl{solar} daily modulations just as well as for sidereal ones, although the distinction can only be made with large statistics.

We have examined in detail the total rate and modulations expected in the CUORICINO and CUORE-0 experiments as well as the planned CUORE setup. Large parts of the parameter space (the WIMP mass, the excited state energy, and the cross-section) can be efficiently explored. While our analysis concentrated on the CUORE setups, we emphasize that this search can be done by other experiments with high-resolution x-ray detectors such as CoGeNT~\cite{Aalseth:2012if}. In fact, a comparison of this measurement between different detectors, e.g. a tellurium-based versus germanium-based, can help to discriminate signal (which is detector \textsl{independent}) from possibly confusing nuclear lines (which are detector \textsl{dependent}).

We also showed that an extra shield, strategically placed, can easily enhance the modulation fraction to $\sim \mathcal{O}(30\%)$. Importantly, while the precise amount by which the modulation can be increased depends on the lifetime, the detailed geometry, the proximity of the extra shield to the detector, and the amount of extra shield used, it can nevertheless be optimized given a particular experimental setup. Finally, we note that in case of a positive detection it will be possible to extract the lifetime with a careful placement of the shield at different distances from the detector.  

\begin{acknowledgments}
IY and MP are  supported in part by funds from the Natural Sciences and Engineering Research Council (NSERC) of Canada. Research at the Perimeter Institute is supported in part by the Government of Canada through Industry Canada, and by the Province of Ontario through the Ministry of Research and Information (MRI). NW is supported by the NSF under grants 0947827 and PHY-1316753.
\end{acknowledgments}

%\newpage
%%\onecolumngrid
\appendix

%%%%%%%%%%%%%%%%
% Appendix A
%%%%%%%%%%%%%%%%

\renewcommand{\theequation}{A-\arabic{equation}}
\setcounter{equation}{0}

\section{Three-level systems}
\label{app:three_level_system}

A concrete field theory realization of the three-level system shown in Fig.~\ref{fig:three-levels} is not hard to find. One example, involving an ${\rm SU}(2)$ dark gauge group, was presented in ref.~\cite{Finkbeiner:2009mi}. The gauge-group is broken at around a GeV and the WIMP, $\chi_i$, which transforms as a triplet of that group receives radiative corrections that split the mass of the three states, $\chi_{1,2,3}$ in such a way as to allow the identification of $\chi_3$ with the ground state, $\chi_2$ with the metastable state, and $\chi_1$ as the excited state. The broken gauge-group is weakly mixed with the Standard Model hypercharge as in the Lagrangian above - Eq.~(\ref{eqn:darkLag}). Therefore, the transition from the metastable state to the excited state through a dark vector-boson exchange against the nucleus is possible. To realize the de-excitation through the magnetic dipole transition, this model should be supplemented with the following interaction terms,
\be
\nonumber
\mathcal{L}_{\rm dipole} &=& \frac{c_1}{\Lambda^2}\epsilon_{ijk} \phi_i \chi_j \sigma_{\mu\nu} F^{\mu\nu} \chi_k \\ &+& \frac{c_2}{\Lambda^2}\epsilon_{ijk} \phi'_i \chi_j \sigma_{\mu\nu} F^{\mu\nu} \chi_k
\ee 
where $c_{1,2}$ are some unknown dimensionless numbers, $\epsilon_{ijk}$ is the antisymmetric tensor of ${\rm SU}(2)$, and the scalar fields $\phi$ and $\phi'$ receive a vacuum expectation value that breaks the dark ${\rm SU}(2)$ completely:  $\langle \phi_3\rangle \ne 0$ and $\langle\phi'_2\rangle \ne 0$. This generates transitions only between $\chi_1 - \chi_2$ and $\chi_1 - \chi_3$, thus maintaining the metastability of $\chi_2$. Higher dimensional operators ultimately induce a transition between $\chi_2-\chi_3$, but the associated rate can be sufficiently slow to allow $\chi_2$ to be metastable on cosmological time-scales.  

The three levels  shown in Fig.~\ref{fig:three-levels} may also be the result of an atomic-like structure for dark matter~\cite{Kaplan:2009de,Kaplan:2011yj}. For example, the \textrm{2S} state in muonic hydrogen is metastable since the direct decay to the \textrm{1S} ground state is restricted by conservation of angular moment to proceed only through a two-photon emission. On the other hand, the \textrm{2S} state enjoys a single photon transition to the excited \textrm{2P} state, which is followed by a rapid decay to the \textrm{1S} ground state.

Finally, this setup can be realized with dark vector-bosons which are weakly mixed with the SM hypercharge as in Eq.\ ~(\ref{eqn:darkLag}). We consider a case where the mass of the vector-boson is such that the decay from the metastable state to the ground states through an on-shell vector-boson is kinematically forbidden. If the mass splitting is smaller than the pair production threshold, $2m_e$,  then the lifetime of this state becomes cosmologically long~\cite{Finkbeiner:2009mi}. The decay from the excited state down to the ground state through an on-shell vector-boson might still be allowed (this requires a tuning in the mass of the vector boson as compared with the energy difference of the three states). This configuration will result in a cosmologically long-lived metastable state which can scatter against the nucleus into the excited state through a dark vector-boson exchange. The excited state then decays down to the ground state through an on-shell dark vector-boson. The dark vector-boson itself is unstable and decays into an electron-positron pair. The lifetime of this decay depends on the kinetic mixing parameter ($\epsilon$ in Eq.~(\ref{eqn:darkLag})), and is given by,
\be
\hspace{-3mm} \Gamma_{V\rightarrow e^+e^-} = \frac{\alpha \epsilon^2}{3} m_V \left(1-\frac{4m_e^2}{m_V^2}\right)^{1/2} \left(1+\frac{2m_e^2}{m_V^2}\right)
\ee
If we consider for example $m_V = 2m_e + 90\keV$ and a transition energy of $2m_e + 100\keV$, then the velocity of the vector boson is about 0.1 c the typical distance it will travel before decay is,
\be
d_{_V} = 20~{\rm cm} \left( \frac{10^{-5}}{\epsilon}\right)^2
\ee
It is beyond the scope of this paper to study the full parameter space of this model, however, the phenomenology is similar to the one described in the paper even though the collision and de-excitation both occur through dark forces and not through a magnetic dipole transition. The main difference is that the de-excitation results in an electron-positron pair which reconstructs the de-excitation energy rather than a single photon.

%However, in our case the relevant quantities are the kinetic energy of the excited DM state after the collision in the shield, $\EX$, and the solid-angle of its direction of travel with respect to the detector, $\Omega$. The angle of scattering in the lab frame with respect to the WIMP's original direction of motion is given in terms of its energy, $\EX$, 
%\be
%\cos\theta_{\rm lab}& =& \frac{\mN}{\mNX}\frac{1}{\sqrt{2\mX \EX v^2}}\Bigg[ \EX  
%\\ \nonumber
% &+& \frac{1}{2}\mNX v^2 \left( \frac{\mX-\mN}{\mN} + \frac{2\delta}{\mX v^2} \right) \Bigg]
%\ee
%The polar angle is uniformly distributed as it is in the COM frame. 
% 
%
%%%%%%%%%%%%%%%%%% section %%%%%%%%%%%%%%%%%%

\bibliography{midm-xray}
\end{document}